\documentclass[preprint,prd,aps,showpacs,nofootinbib]{revtex4}
\usepackage{inputenc}
\usepackage{epsfig,float}
\usepackage{amsfonts}
\usepackage{amsmath}

\input{epsf}
\newcommand{\be}{\begin{eqnarray}}
\newcommand{\ee}{\end{eqnarray}}
\newcommand{\ba}{\begin{array}}
\newcommand{\ea}{\end{array}}

\restylefloat{figure}
\restylefloat{table}

\begin{document}

\title{Accessing baryon to meson transition distribution amplitudes in meson production in association with a high invariant
mass lepton pair at GSI-FAIR with $\mathbf{\bar{P}}$ANDA}

\author{J.~P.~Lansberg$^1$, B.~Pire$^2$,  K.~Semenov-Tian-Shansky$^{3}$, L.~Szymanowski$^{4}$ }
\affiliation{
$^1$ IPNO,   Universit\'{e} Paris-Sud, CNRS/IN2P3, 91406 Orsay, France \\
$^2$ CPhT, \'{E}cole Polytechnique, CNRS,  91128 Palaiseau, France  \\
$^3$ IFPA, d\'{e}partement AGO,  Universit\'{e} de  Li\`{e}ge, 4000 Li\`{e}ge,  Belgium \\
$^4$ National Center for Nuclear Research (NCBJ), 00-681 Warsaw, Poland
}

\preprint{CPHT-RR072.1012}

\begin{abstract}
Nucleon-antinucleon annihilation into a near backward- (or forward-) produced meson
and a high invariant mass lepton pair admits a factorized description in terms of
antinucleon (or nucleon) distribution amplitudes (DAs) and nucleon to meson (or antinucleon to meson)
transition distribution amplitudes (TDAs). We estimate the cross section of
backward (and forward) pion and $\eta$-meson production in association with a high invariant mass
lepton pair for the kinematical conditions of GSI-FAIR. The cross sections are found to be large
enough to be measured with the ${\bar{\rm P}}$ANDA detector. Interesting phenomenological applications of the approach
are thus expected.

\end{abstract}

\maketitle
\thispagestyle{empty}
\renewcommand{\thesection}{\arabic{section}}
\renewcommand{\thesubsection}{\arabic{subsection}}

\section{Introduction}
\label{Sec-1}

The ambitious experimental program for the ${\bar{\rm P}}$ANDA facility at GSI-FAIR
(see  {\cite{Lutz:2009ff,Wiedner:2011mf})
will provide the experimental access to new classes of hard reactions
for which the factorized description applies.
Study of these reactions will open a new eyehole to look on hadron's interior and will bring
a new perspective to the problem of hadronic structure  in terms of the fundamental degrees of freedom of QCD.

A tempting possibility is to consider nucleon-antinucleon  annihilation into a lepton pair with
a high invariant mass
$q^2 \equiv Q^2$
in association with a light meson
${\mathcal{M}}=\{\pi,\,\eta,\, \rho, \omega,\, ... \}$:
\be
\bar{N} (p_{\bar{N}}) N (p_N) \rightarrow \gamma^*(q) {\mathcal{M}}(p_{\mathcal{M}}) \rightarrow \ell^+(p_{\ell^+})
\ell^-(p_{\ell^-}) {\mathcal{M}}(p_{\mathcal{M}}).
\label{BarNNannihilation reaction}
\ee
The analysis of
\cite{Pire:2005ax,Lansberg:2007se},
identifies two similar factorization regimes for the reaction
(\ref{BarNNannihilation reaction})
corresponding to forward and backward peaks of the meson production cross
section.

Let us now specify our conventions%
\footnote{Definitions of $t$ and $u$ here match Ref.~\cite{Lansberg:2011aa}
and differ from those of Ref.~\cite{Lansberg:2007se}}. The $z$-axis is chosen along the colliding nucleon-antinucleon.
For the ${\bar{\rm P}}$ANDA setup it is natural to select as the positive direction of $z$  the direction
the antinucleon is moving in the
$\bar{N} N$
center-of-mass (CMS) frame.
\begin{itemize}
\item Within these conventions, the kinematical regime in which the $u$-channel momentum transfer squared,
$ u  \equiv  (p_N-p_{\mathcal{M}})^2 $,
is small  corresponds to the meson moving in the direction of the initial nucleon that is the
\emph{backward}
direction.
\item Analogously, the kinematical regime in which the $t$-channel momentum transfer squared,
$ t  \equiv  (p_{\bar{N}}-p_{\mathcal{M}})^2 $,
is small, corresponds to the meson
moving in the direction of the initial antinucleon that is the
\emph{forward}
direction.
\end{itemize}
A detailed account of the corresponding kinematics is presented in 
Appendix~\ref{App_A}.

The backward  and forward factorization regimes for the reaction
(\ref{BarNNannihilation reaction})
are casted with the help of two
collinear factorization theorems  schematically depicted on
Fig.~\ref{Fig1}.
\begin{enumerate}
\item  For backward kinematics the collinear factorization theorem
(see the left panel of
Fig.~\ref{Fig1})
is valid once
$s=(p_N+p_{\bar{N}})^2 \equiv W^2$
and the invariant mass of the lepton pair
$q^2 \equiv Q^2$
are large;
the corresponding skewness variable $\xi^u$ (\ref{def_xiu}),
which characterizes the $u$-channel longitudinal momentum transfer,
is fixed;
$|u|$
is small as compared to
$Q^2$
and
$W^2$.
The meson $\cal M$ is produced in the near backward direction in the
$\bar{N}N$
CMS. We refer to this
kinematics as the backward, or the $u$-channel, factorization regime.

\item
Analogously, the $t$-channel collinear factorization theorem  presented on the right panel of
Fig.~\ref{Fig1} is valid
once  $W^2$ and $Q^2$
are large;
the corresponding skewness variable
$\xi^t$ (\ref{defxit})
is fixed;
$|t|$
is small as compared to
$Q^2$
and
$W^2$.
The meson $\cal M$ is produced in the near forward direction in
$\bar{N}N$ CMS, and we refer to this
kinematics as the forward, or the $t$-channel, factorization regime.
\end{enumerate}
It is worth emphasizing that for the moment the rigorous proof of the collinear factorization
theorems for meson production in association with a high invariant
mass lepton pair in $N  \bar{N}$ annihilation as well as for backward meson electroproduction does not exist.
Some physical arguments in favor of this kind of factorization for the spacelike version of the reaction
(\ref{React_PANDA})
were presented in
Refs.~\cite{Frankfurt:1999fp, Frankfurt:2002kz}.
In this paper we do not aim to prove the relevant factorization theorems but rather perform
the feasibility estimates to try to see some evidences of the factorized description in the experiment.

\begin{figure}[h]
 \begin{center}
 \epsfig{figure= 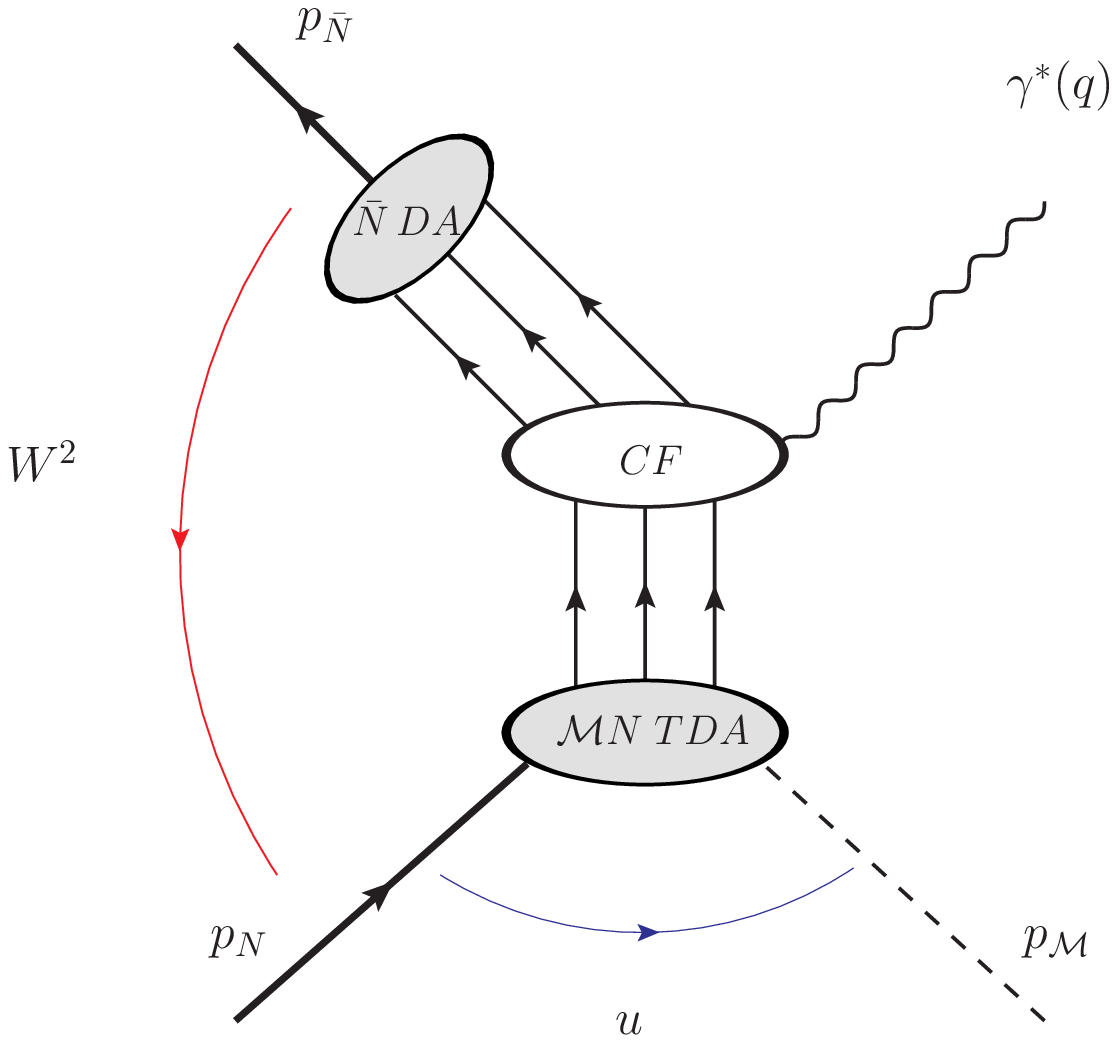 , height=5.8cm}
 \epsfig{figure=  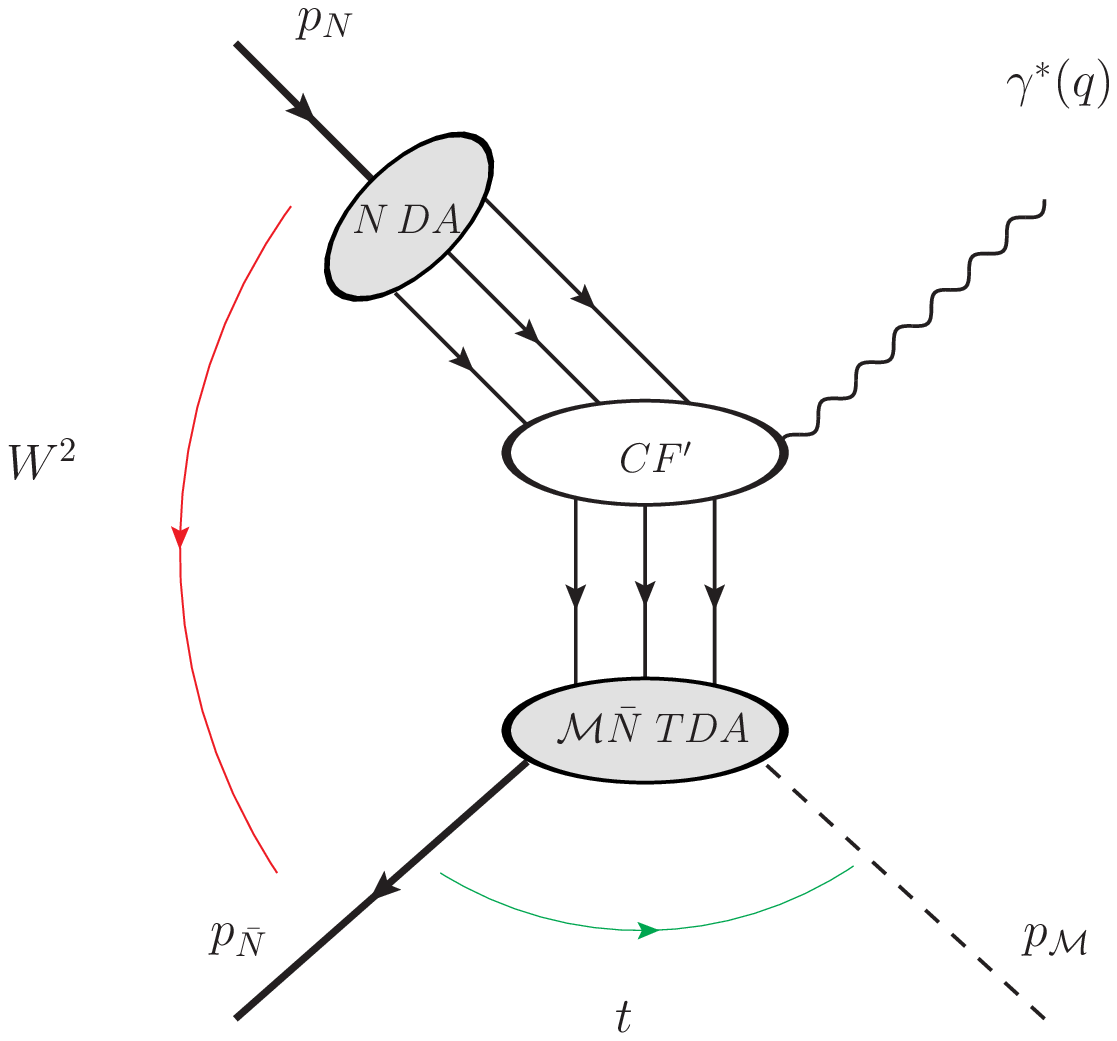 , height=5.8cm} \ \ \
   \end{center}
     \caption{ Two possibilities for  collinear factorization  of the annihilation
     process $N \bar{N} \to \gamma^*(q) \mathcal{M}(p_\mathcal{M})$. {\bf Left panel:}  backward  kinematics ($|u| \sim 0$) .
           {\bf Right panel:}
     forward kinematics ($|t| \sim 0$).  $\bar{N}(N)$ DA stands for the distribution amplitude
     of antinucleon (nucleon); $\mathcal{M} N (\mathcal{M} \bar{N})$ TDA stands for the transition
     distribution amplitude from a nucleon (antinucleon) to a meson;
      CF and CF' denote hard subprocess amplitudes (coefficient functions).  }
\label{Fig1}

\end{figure}

The factorization theorems  for the reaction
(\ref{BarNNannihilation reaction})
presented in Fig.~\ref{Fig1}
involve two kinds of nonperturbative objects:
the conventional antibaryon (or  baryon) distribution amplitudes (DAs)
and the baryon to meson (or antibaryon to meson) transition distribution amplitudes (TDAs).
The baryon to meson (antibaryon to meson) TDAs are defined through
baryon (antibaryon)-meson matrix elements of
the nonlocal three quark (antiquark) operator on the light cone
\cite{Radyushkin:1977gp,Efremov:1978rn,Lepage:1980,Chernyak:1983ej,Chernyak_Nucleon_wave}:
\be
&&
\hat{O}^{\alpha \beta \gamma}_{\rho \tau \chi}( \lambda_1 n,\, \lambda_2 n, \, \lambda_3 n)
 =
\varepsilon_{c_{1} c_{2} c_{3}}
\Psi^{c_1 \alpha}_\rho(\lambda_1 n)
\Psi^{c_2 \beta}_\tau(\lambda_2 n)
\Psi^{c_3 \gamma}_\chi (\lambda_3 n)
;
\nonumber \\ &&
\hat{\bar{O}}_{\alpha \beta \gamma \; \rho \tau \chi}( \lambda_1 n,\, \lambda_2 n, \, \lambda_3 n)
 =
\varepsilon_{c_{1} c_{2} c_{3}}
\bar{\Psi}^{c_1 }_{\alpha \;\rho}( \lambda_1 n)
\bar{\Psi}^{c_2 }_{\beta \; \tau}(  \lambda_2 n)
\bar{\Psi}^{c_3 }_{\gamma \;  \chi} (\lambda_3 n)
.
\label{operators}
\ee
Here
$\alpha$, $\beta$, $\gamma$
stand for quark (antiquark) flavor indices and
$\rho$, $\tau$, $\chi$
denote the Dirac spinor indices. Antisymmetrization stands over the color group indices
$c_{1,2,3}$%
\footnote{See App.~\ref{App_B1} for the details of our index conventions. }.
The gauge links in
(\ref{operators}) can be omitted in  the light-like gauge
$A^+=0$. For the case when the meson
${\mathcal{M}}$
is a pseudoscalar, to the leading twist-$3$ accuracy,
the form factor decomposition of
${\mathcal{M}}N$
(${\mathcal{M}}\bar{N}$)
matrix elements of (\ref{operators})
involves
eight
invariant
${\mathcal{M}} N$ (${\mathcal{M}} \bar{N}$)
TDAs
$H^{{\mathcal{M}} N}$ ($H^{{\mathcal{M}} \bar{N}}$),
each being a function of the three longitudinal momentum fractions
$x_i$ ($\sum_{i} x_i =2 \xi$),
of the corresponding skewness variable
$\xi$
and of the momentum transfer squared
$\Delta^2$,
\be
&&
H^{({\mathcal{M}}N), \, ({\mathcal{M}} \bar{N})}  (x_1,x_2,x_3,\xi^{u,t},{\Delta^{u,t}}^{2})
\nonumber \\ &&
 \equiv
\{V_{1,2}^{({\mathcal{M}}N), \, ({\mathcal{M}} \bar{N})}, A_{1,2}^{(MN), \, (M \bar{N})} ,T_{1,2,3,4}^{({\mathcal{M}}N), \, ({\mathcal{M}} \bar{N})}   \} (x_1,x_2,x_3,\xi^{u,t},{\Delta^{u,t}}^{2}).
\label{TDAsScalar}
\ee
The isotopic symmetry and the charge conjugation invariance further reduce the number of
independent  nucleon to meson TDAs. In particular, $C$-invariance allows us to express all
${\mathcal{M}} \bar{N}$
TDAs through
${\mathcal{M}} N$
TDAs (see App.~\ref{App_B}).

Extensive studies of the properties and of the physical interpretation of $\pi N$ TDAs are presented in
Refs. \cite{Pire:2005ax,Lansberg:2007ec,Lansberg:2007se,Pire:2010if,Pire:2011xv,Lansberg:2011aa}.
Conceptually, the
${\mathcal{M}} N$
TDAs share many common features both with the GPDs and the nucleon DAs.
Indeed, the crossing transformation relates ${\mathcal{M}} N$ TDAs with
the $\bar{{\mathcal{M}} }N$
generalized distribution amplitudes (GDAs), defined as the matrix element of the same
light cone operator between
$\bar{{\mathcal{M}}} N$
state and the vacuum. In the pion case, this allows to establish useful constraints from the chiral
dynamics since $\pi N$ GDAs reduce to combinations of the usual nucleon DAs in the soft pion limit.
In addition,
$\pi N$
TDAs were recently estimated  within the light-cone quark model
\cite{Pasquini:2009ki}.
The hadronic
matrix elements of QCD operators are extensively studied on the lattice. In particular,
$\pi N$
matrix elements of the operators corresponding to the Mellin moments of
$\pi N$ TDAs
have been recently considered on the lattice
by several groups, see
\cite{Aoki:1999tw},
\cite{Aoki:2006ib}
and Refs. therein.

On the other hand, similarly to the GPD case
\cite{Burkardt:2000za,Burkardt:2002ks,Diehl:2002he,Pir},
a comprehensible physical picture  may be obtained by Fourier transforming ${\cal M} N$ TDAs to the impact parameter space.

\section{Modelling $\pi N$ TDAs}
\label{Sec-2}

First estimates of the cross section of
 $\pi^0$
 production in association with a high-$Q^2$ lepton pair in
 $\bar{p} p$
annihilation for ${\bar{\rm P}}$ANDA@GSI-FAIR conditions within the factorized description involving
$\pi N$
TDAs were presented in
\cite{Lansberg:2007se}.
In this analysis, an oversimplified  model for $\pi N$ TDAs based on the soft pion theorem was employed.
This model has obvious drawbacks: its domain of applicability is limited to the immediate vicinity
of the soft-pion threshold; TDAs in this model do not have any intrinsic $\Delta_T^2$ dependence and
finally, the cross channel leading baryon exchange contribution, which turns
out to be dominant in many cases, was not taken into account.

In Ref.~\cite{Lansberg:2011aa}, a two component model for $\pi N$ TDAs was proposed.
It includes the spectral part casted in terms of quadruple distributions and a $D$-term
like contribution which is determined by the
nucleon pole exchange in the cross channel (see Fig.~\ref{Fig_Nucleon_pole}). Quadruple distributions are
fixed with the help of the soft pion theorem for
$\pi N$
TDAs in terms of nucleon DAs. Once the pion mass is neglected
($m=0$)
in the strict soft pion limit
($\xi=1$, $\Delta^2=M^2$),
this model gives the same predictions for the
$N \bar{N} \rightarrow \ell^+ \ell^- \pi $
cross section as the model employed in
\cite{Lansberg:2007se}.

\begin{figure}[H]
 \begin{center}
   \epsfig{figure= 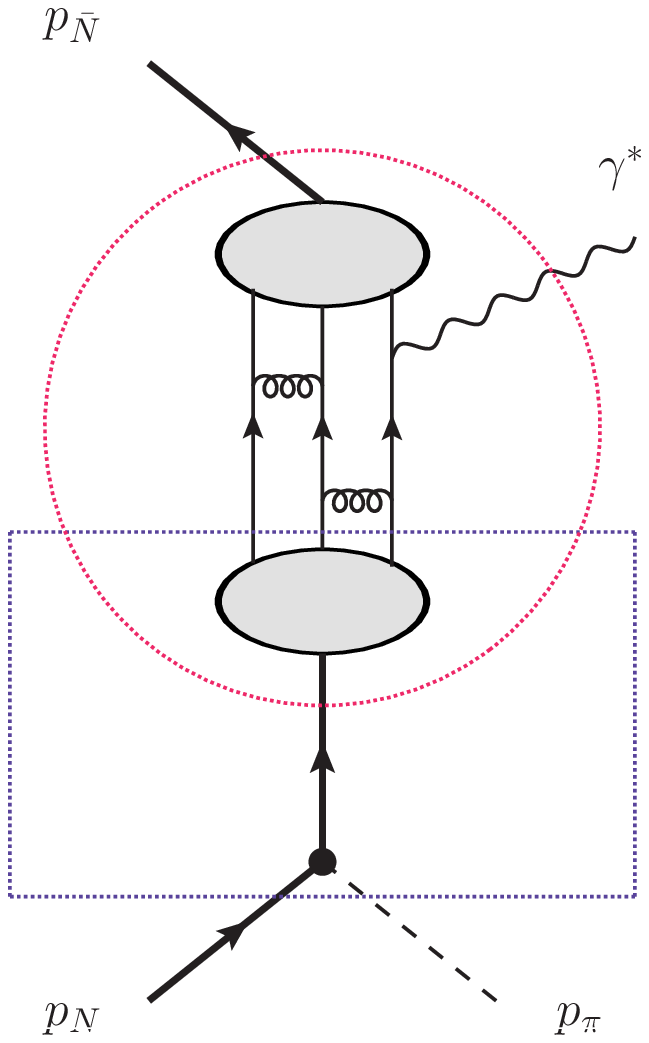 , height=4.8cm} \ \ \ \ \ \ \ \
    \epsfig{figure= 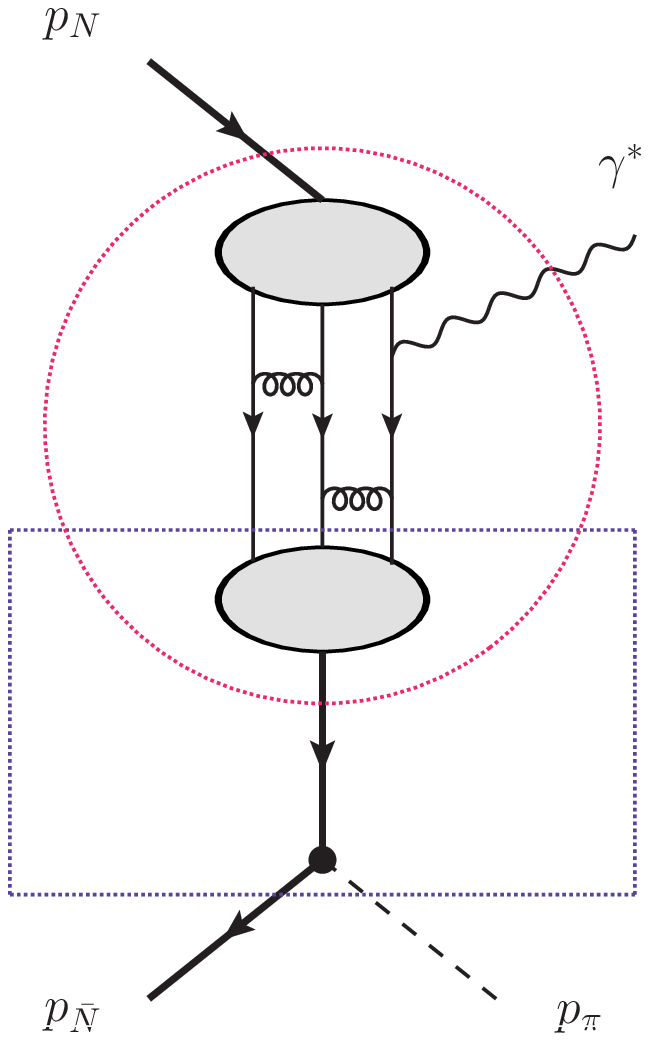 , height=4.8cm}
     \caption{Nucleon pole exchange model for $\pi N$ (left) and $\pi \bar{N}$ (right) TDAs;
dashed circles contain typical LO graph for the nucleon (antinucleon) electromagnetic form factor
in perturbative QCD; rectangles contain the nucleon pole contribution into $\pi N$ ($\pi \bar{N}$) TDAs.
      }
\label{Fig_Nucleon_pole}
\end{center}
\end{figure}

As it was found out in
\cite{Lansberg:2011aa},
the contribution of the spectral part dominates the cross section for large values of
$\xi \approx 1$
(which is difficult to reach in  experiments), while for small and intermediate values of
$\xi$,
the  nucleon pole exchange part provides the dominant contribution to the
$N \bar{N} \rightarrow \gamma^* \pi$
cross section.

In Sec.~\ref{Ssec-41}, we present our estimates of
$N \bar{N} \rightarrow \ell^+ \ell^- \pi$
cross section within our factorization scheme in the $u$-channel factorization regime%
\footnote{In the remainder of the paper, in most cases we  omit superscripts $u$ and $t$ for the kinematical quantities
($\xi$, $\Delta^2$, $p$, $n$ {\it etc.})
referring to the factorization regime. }
 (near backward pion production)
using the nucleon pole exchange model for
$\pi N$ TDAs.

The explicit expression for the contribution of the cross-channel nucleon exchange
into the isospin-$\frac{1}{2}$
$\pi N$
TDAs was obtained in \cite{Pire:2011xv},
\be
&&
\big\{ V_1, \, A_1 , \, T_1  \big\}^{(\pi^0 p)} ( 
x_i, \xi,\Delta^2)\Big|_{N(940)}
\nonumber \\ &&
 =\Theta_{\rm ERBL}(x_1,x_2,x_3) \times  (g_{\pi NN}) \frac{M f_\pi}{\Delta^2-M^2}    \frac{1}{(2 \xi) }
 \big\{ V^p,\,A^p, \,T^p  \big\}\left( \frac{x_1}{2 \xi}, \frac{x_2}{2 \xi}, \frac{x_3}{2 \xi} \right);
  \nonumber \\ &&
\big\{ V_2, \, A_2 , \, T_2  \big\}^{(\pi^0 p)} (x_i ,\xi,\Delta^2)\Big|_{N(940)}=
\frac{1}{2}
\big\{ V_1, \, A_1 , \, T_1  \big\}^{(\pi^0 p)} (x_i,\xi,\Delta^2)\Big|_{N(940)};   \nonumber \\ &&
\big\{  T_3, \, T_4  \big\}^{(\pi^0 p)} (x_i,\xi,\Delta^2)\Big|_{N(940)}=0,
\label{Nucleon_exchange_contr_VAT}
\ee
where
$V^p$, $A^p$ and $T^p$
stand for the nucleon DAs;
$g_{\pi NN}\approx 13$
is the pion-nucleon phenomenological coupling and
\be
\Theta_{\rm ERBL}(x_1,x_2,x_3)  \equiv  \prod_{k=1}^3 \theta(0 \le x_k \le 2 \xi)
\label{theta_ERBL}
\ee
ensures the pure Efremov-Radyushkin-Brodsky-Lepage (ERBL) support.

For
$\pi^-n$ TDAs,
we get
\be
&&
\big\{ V_{1,2}, \, A_{1,2} , \, T_{1,2}  \big\}^{(\pi^- n)}(x_i,\xi,\Delta^2)\Big|_{N(940)}=
\sqrt{2}  \big\{ V_{1,2}, \, A_{1,2} , \, T_{1,2}  \big\}^{(\pi^0 p)}(x_i,\xi,\Delta^2)\Big|_{N(940)};
\nonumber \\ &&
\big\{  T_3, \, T_4  \big\}^{(\pi^- n)} (x_i,\xi,\Delta^2)\Big|_{N(940)}=0.
\ee

We also present some estimates of the $N \bar{N} \rightarrow \ell^+ \ell^- \pi$ cross section within the two component model
for $\pi N$ TDAs of Ref.~\cite{Lansberg:2011aa}.

\section{Cross section of $\bar{N}N \to \gamma^* \pi \to \pi \ell^+ \ell^-$ }
\label{Sec-3}

Below we review the formulas for the cross section of
\be
N(p_N,s_N)+\bar{N}(p_{\bar{N}},s_{\bar{N}})
\rightarrow
\gamma^*(q)+\pi(p_\pi)
\rightarrow
\ell^+(p_{\ell^+})+\ell^-(p_{\ell^-})+\pi(p_\pi)
\label{React_PANDA}
\ee
established in
\cite{Lansberg:2007se}.
The starting point is the general formula for the unpolarized differential cross section of the reaction
(\ref{React_PANDA})
(see {\it} e.g. \cite{CORE}),
\be
d \sigma = \frac{1}{2 (2 \pi)^5
{\Lambda(W^2,M^2,M^2)}} |\overline{\mathcal{M}^{ N \bar{N} \rightarrow \ell^+ \ell^- \pi}}|^2
d_3({\rm LIPS}),
\ee
where
$\Lambda(x,y,z)= \sqrt{x^2+y^2+z^2-2xy-2xz-2yz}$
is the usual Mandelstam function.

 Following the standard procedure, the $3$-particle
differential Lorentz invariant phase space
$d_3({\rm LIPS})$
is decomposed into two $2$-particle phase subspaces: those of
$\gamma^* \pi$
and
$\ell^+ \ell^-$
systems. The former can be easily computed in
$\bar{N} N$
CMS while the latter is computed in
$\ell^+ \ell^-$
CMS yielding the result,
\be
d \sigma = \frac{1}{2 (2 \pi)^5
{\Lambda(W^2,M^2,M^2)}}
|\overline{\mathcal{M}^{ N \bar{N} \rightarrow \ell^+ \ell^- \pi }}|^2
\frac{d \Omega^*_\pi}{8 W^2 }
{\Lambda(W^2,Q^2,m^2)} \frac{d \Omega_\ell}{8},
\ee
where
$d\Omega^*_\pi \equiv  d \cos \theta_\pi^* d \varphi_\pi^*$
in $N \bar{N}$ CMS.
By $d \Omega_\ell
\equiv
d \cos \theta_\ell d \varphi_\ell$,
we denote the produced lepton solid angle in $\ell^+ \ell^-$ CMS. By expressing
$ \cos \theta_\pi^*$
through
$u = (p_{N}-p_\pi)^2$,
\be
du = \frac{d \cos \theta_\pi^*}{2 W^2}
{\Lambda(W^2,M^2,M^2)}
{\Lambda(W^2,Q^2,m^2)}
\ee
and  integrating over the azimuthal angle
$\varphi_\pi^*$
of the produced pion and over the  azimuthal angle of the lepton pair
$\varphi_\ell$,
the following formula for the differential cross section of the reaction
(\ref{React_PANDA})
is established:
\be
\frac{d \sigma}{du d Q^2 d \cos \theta_{\ell}}=
 \frac{\int d \varphi_\ell |\overline{\mathcal{M}^{N \bar{N} \rightarrow \ell^+ \ell^- \pi}}|^2 }
 {64 W^2 (W^2-4M^2) (2 \pi)^4}.
\ee

The next step is to express the average-squared amplitude
$|\overline{\mathcal{M}^{ N \bar{N} \rightarrow \ell^+ \ell^- \pi }}|^2$
through the amplitude
$\mathcal{M}_{\lambda}^{s_N s_{\bar{N}}}$
of
$N \bar{N} \rightarrow \gamma^* \pi $.
Within the factorized approach of
\cite{Lansberg:2007se},
at leading order in
$\alpha_s$,
the amplitude
$\mathcal{M}_{\lambda}^{s_N s_{\bar{N}}}$
reads
\be
&&
\mathcal{M}_{\lambda}^{s_N s_{\bar{N}}}=
\mathcal{C} \frac{1}{Q^4}
\Big[
\mathcal{S}_{\lambda}^{s_N s_{\bar{N}}}
\mathcal{I}(\xi, \Delta^2)
-
\mathcal{S'}_{\lambda}^{s_N s_{\bar{N}}}
\mathcal{I}'(\xi, \Delta^2)
\Big],
\label{Def_ampl_M}
\ee
where
\be
{ \cal C}=-i \frac{(4 \pi \alpha_s)^2 \sqrt{4 \pi \alpha_{em}} f_N^2}{54 f_\pi} .
\label{Def_CalC}
\ee
Here,
$f_\pi=93$
MeV is the pion weak decay constant and
$f_N $
is a constant, which determines the value of the dimensional nucleon wave
function at the origin
\cite{Chernyak_Nucleon_wave};
$\alpha_{em}\simeq \frac{1}{137}$
is the electromagnetic fine structure constant and
$\alpha_s \simeq  \frac{1}{3}$
is the strong coupling constant.

The spin structures
$\mathcal{S}$ and $\mathcal{S'}$
are defined as
\be
&&
\mathcal{S}_{\lambda}^{s_N s_{\bar{N}}} \equiv
\bar{V}(p_{\bar{N}},s_{\bar{N}}) \hat{\epsilon}^*(\lambda) \gamma_5 U(p_N,s_N);
\nonumber \\ &&
\mathcal{S'}_{\lambda}^{s_N s_{\bar{N}}} \equiv
\frac{1}{M}
\bar{V}(p_{\bar{N}},s_{\bar{N}}) \hat{\epsilon}^*(\lambda) \hat{\Delta}_T \gamma_5 U(p_N,s_N).
\ee
We employ the standard ``hat'' notation for the contraction of a four-vector with the Dirac matrices.
$\epsilon(\lambda)$
stands for the polarization vector of the virtual photon.
$\mathcal{I}$
and
$\mathcal{I'}$
denote the convolution integrals of
$\pi N$
TDAs and antinucleon DAs with the hard scattering kernels computed from the set of
$21$
relevant scattering diagrams (see \cite{Lansberg:2007ec}):
\be
&&
\{\mathcal{I}, \mathcal{I}'\}(\xi,\Delta^2) \equiv { {\int^{1+\xi}_{-1+\xi} }\! \! \!
dx_1  {\int^{1+\xi}_{-1+\xi} }\! \! \! dx_2  {\int^{1+\xi}_{-1+\xi} }\! \! \!dx_3 \, \delta(x_1+x_2+x_3-2\xi)
}
\nonumber \\ &&
\times
{{\int^{1}_{0} }\! \! \! dy_1  {\int^{1}_{0} }\! \! \! dy_2  {\int^{1}_{0} }\! \! \!dy_3 \, \delta(y_1+y_2+y_3-1)}
{\Bigg(2\sum_{\alpha=1}^{7} \{ R_{\alpha}, R_{\alpha}'\}+
\sum\limits_{\alpha=8}^{14} \{ R_{\alpha},  R_{\alpha}'\} \Bigg)}.
\label{Def_IandIprime}
\ee
The integrals  in
$x_i$'s ($y_i$'s)
stand  over the support of
$\pi N$
TDA (antinucleon DA).

For the $u$-channel factorization regime of
$p \bar{p} \rightarrow \pi^0 \ell^+ \ell^-$,
the coefficients
$R_\alpha$
and
$R'_\alpha$ ($\alpha=1,..,14$)
correspond to the coefficients
$T_\alpha$
and
$T'_\alpha$
presented in  Table~I of
\cite{Lansberg:2007ec}
up to an overall irrelevant phase factor
${{\eta_N^*} }^{-1} \eta_q^{-3}$
and to the replacement
$-i \varepsilon \rightarrow  i \varepsilon$
in the denominators of hard scattering kernels. This latter change mirrors the fact that we consider
$\gamma^*$
in the final state rather than in the initial state as it is for backward pion electroproduction.

We also note that   Ref.~\cite{Lansberg:2007ec} employed a somewhat
inadequate parametrization of
$\pi N$
TDAs, which does not maintain the polynomiality property in its simple form.
In this paper, we use the parametrization of
\cite{Pire:2011xv} which maintains the property of polynomiality.
The relation between the two parametrizations is presented in App.~A  of
\cite{Lansberg:2011aa}.
For
$n \bar{p} \rightarrow \pi^- \ell^+ \ell^-$
reaction the coefficients
$R_\alpha$
and
$R'_\alpha$
are the same as for
$p \bar{p} \rightarrow \pi^0 \ell^+ \ell^-$
channel up to an obvious change
\be
\{V_{1,2}, \, A_{1,2}, \, T_{1,2,3,4}\}^{( \pi^0 p)} \rightarrow \{V_{1,2}, \, A_{1,2}, \, T_{1,2,3,4}\}^{ ( \pi^- n)}.
\ee

The differential cross section for unpolarized beam and unpolarized target is computed
from the squared amplitude
(\ref{Def_ampl_M})
averaged over spins of the initial particles:
\be
|\overline{\mathcal{M}_{\lambda \lambda'}}|^2= \frac{1}{4} \sum_{s_N \, s_{\bar{N}}}
\mathcal{M}_{\lambda}^{s_N s_{\bar{N}}}
\left( \mathcal{M}_{\lambda'}^{s_N s_{\bar{N}}} \right)^*
\ee
with the given combination of photon helicities
$\lambda$, $\lambda'=T,L$.
$|\overline{\mathcal{M}_{L L}}|^2$
$|\overline{\mathcal{M}_{L T}}|^2$
and
$|\overline{\mathcal{M}_{T L}}|^2$
vanish   at the leading twist accuracy.
We employ the relation
\be
\sum_{\lambda_T}  \epsilon^\nu(\lambda) \epsilon^{\mu *}(\lambda)
=-g^{\mu \nu}+\frac{1}{(p \cdot n)}(p^\mu n^\nu+p^\nu n^\mu)
\ee
to sum over the transverse polarizations of the virtual photon and we get
\be
|\overline{\mathcal{M}_{T}}|^2 \equiv \sum_{\lambda_T} |\overline{\mathcal{M}_{T T}}|^2 =
\frac{1}{4} |\mathcal{C}|^2 \frac{2(1+\xi)}{\xi Q^6} \big( |\mathcal{I}|^2- \frac{\Delta_T^2}{M^2} |\mathcal{I}'|^2 \big),
\ee
where
$\mathcal{C}$
is defined in
(\ref{Def_CalC}).

The averaged-squared amplitude for the process
(\ref{React_PANDA})
reads 
\be
|\overline{\mathcal{M}^{ N \bar{N} \rightarrow \ell^+ \ell^- \pi }}|^2=
\frac{1}{4}
\sum_{s_p, \, s_{\bar{p}}, \, \lambda, \, \lambda'}
\mathcal{M}_{\lambda}^{s_p s_{\bar{p}}}
\frac{1}{Q^2}
e^2 {\rm Tr}
\left\{
\hat{p}_{\ell^-} \hat{\epsilon}(\lambda) \hat{p}_{\ell^+} \hat{\epsilon}^*(\lambda')
\right\}
\frac{1}{Q^2}
\left( \mathcal{M}_{\lambda'}^{s_p s_{\bar{p}}} \right)^*.
\label{WF_Cross_sec}
\ee
As pointed above, at the leading twist accuracy, only the transverse polarization of the virtual photon
is contributing. After computing the relevant trace in
(\ref{WF_Cross_sec})
in the
$\ell^+ \ell^-$
CMS and integrating over the lepton polar angle
$\varphi_\ell$,
we finally get:
\be
 \int d \varphi_{\ell} \, |\overline{\mathcal{M}^{ N \bar{N} \rightarrow \ell^+ \ell^- \pi }}|^2
\Big|_{\rm Leading \, twist}=
|\overline{\mathcal{M}_{T}}|^2 \frac{2 \pi e^2(1+\cos^2 \theta_\ell)}{Q^2}.
\ee

A kinematical cut in
$\Delta_T^2$,
that is equivalent to a cut in
$\cos \theta_\pi^*$ (\ref{CosThetaPi_tregime}),
is a convenient way to select near backward and near forward kinematical configurations.
On Fig.~\ref{FigAngls}
with solid lines, we show the dependence of the CMS scattering angles $\theta^*_\pi$ (see eqs.
(\ref{CosThetaPi_tregime}) , (\ref{Cos_theta_u}))
for the $u$-channel (left panel) and the $t$-channel (right panel) factorization regimes as the function of ${\Delta_T^{u}}^2_{\min} $ and
${\Delta_T^{t}}^2_{\min}$ respectively, for $W^2=5$ GeV$^2$, $\xi=0.3$.
For these kinematical parameters, the ${\Delta_T^{u}}^2_{\min}=-0.2$ GeV$^2$ cut corresponds to selecting pions flying backward into
a cone with the aperture of $\sim 120^{\rm o}$ in the $\bar{N} N$ CMS.

In order to implement the effect of the
$\Delta_T^2$
cut,  we consider the cross section
integrated over the appropriate bin in
$u$.
Indeed,
$u$
can be expressed as follows as the function of
$\xi$
and
$\Delta_T^2$:
\be
u=\frac{\Delta_{T}^2 (\xi +1)}{1-\xi }+\frac{2 \xi  \left(M^2 (1-\xi )-m^2 (\xi +1)\right)}{1-\xi ^2}.
\ee
Therefore, the bin
${\Delta_{T}^2}_{\min} \le \Delta_T^2 \le 0 $
corresponds to the following bin in
$u$:
$u_{\min} \le u \le u_{\max}$,
where
\be
&&
u_{\min}(\xi,{\Delta_{T}^2}_{\min} )=\frac{{\Delta_{T}^2}_{\min} (\xi +1)}{1-\xi }+\frac{2 \xi  \left(M^2 (1-\xi )-m^2 (\xi +1)\right)}{1-\xi ^2};
\nonumber
\\
&&
u_{\max}(\xi)=\frac{2 \xi  \left(M^2 (1-\xi )-m^2 (\xi +1)\right)}{1-\xi ^2}.
\ee

\begin{figure}[H]
 \begin{center}
\epsfig{figure= 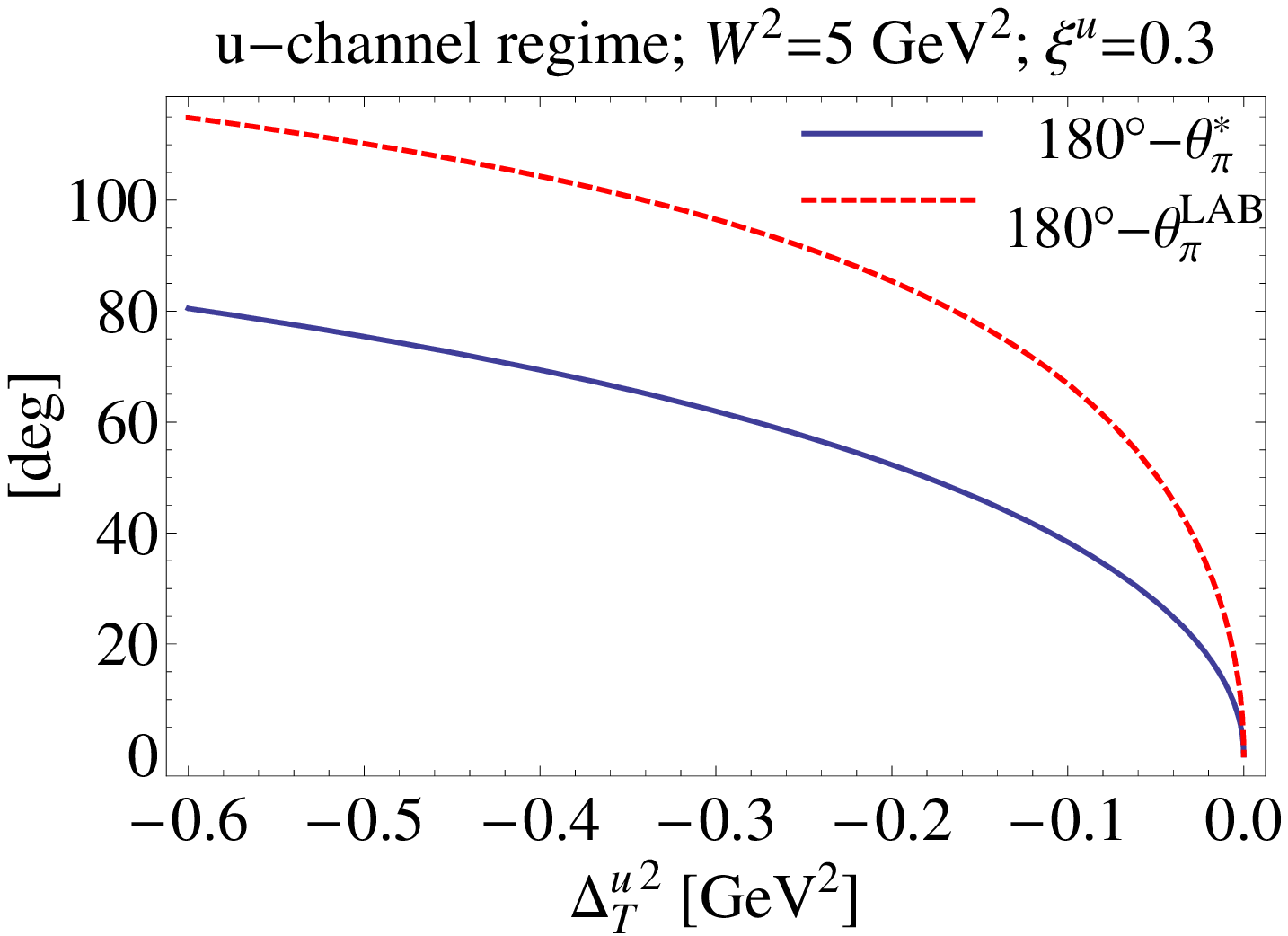 , height=5.2cm}
 \epsfig{figure=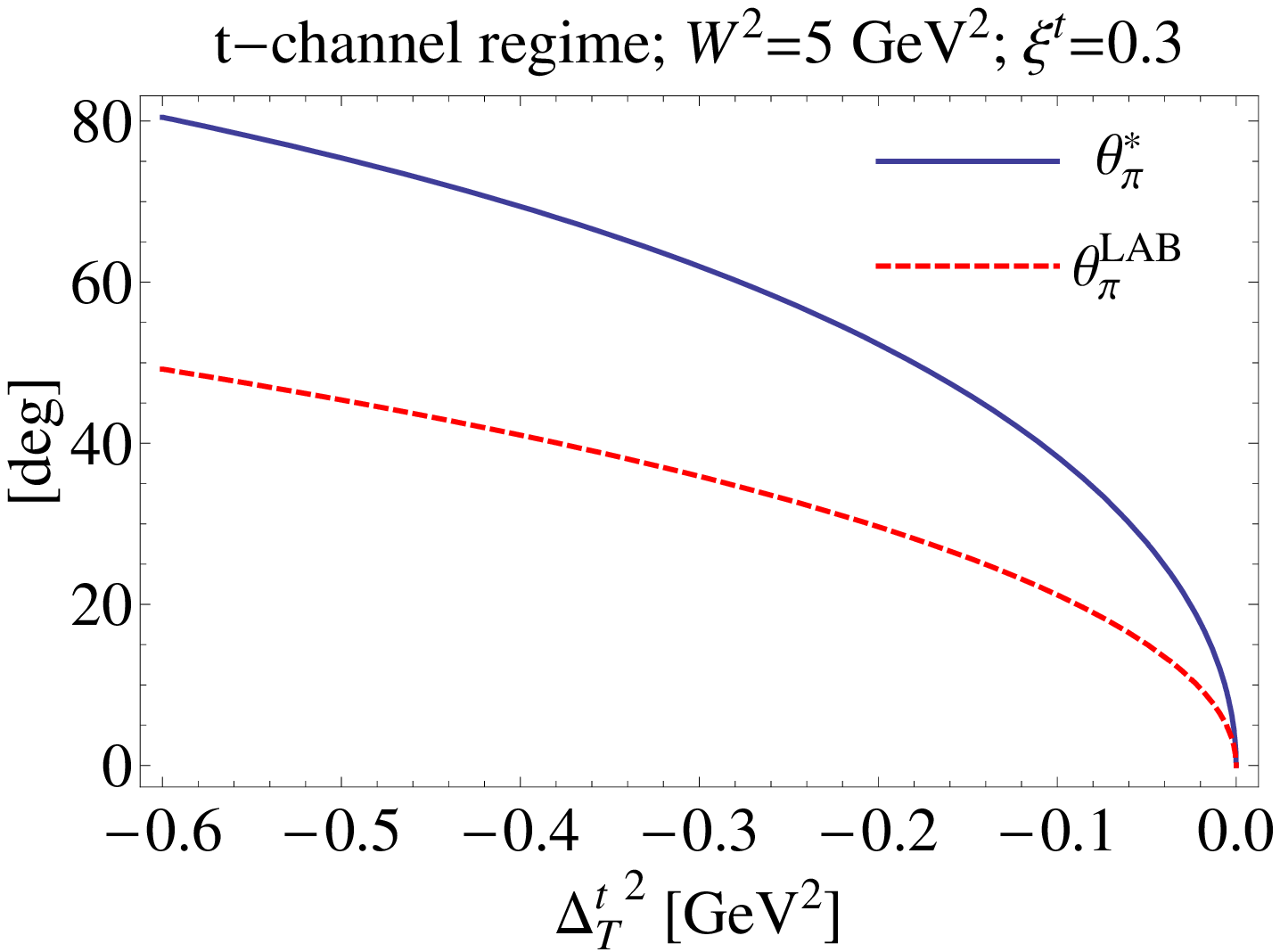 , height=5.2cm}
     \caption{The solid lines illustrate the dependence of the CMS scattering angles $\theta^*_\pi$
for the $u$-channel (left panel) and the $t$-channel (right panel) factorization regimes for the reaction
(\ref{React_PANDA})
as  functions of ${\Delta_T^{u}}^2_{\min} $ and
${\Delta_T^{t}}^2_{\min} $, respectively. The dashed lines illustrate the dependence of the LAB frame  scattering angles $\theta^{\rm LAB}_\pi$
for the two factorization regimes as the function of ${\Delta_T^{u}}^2_{\min}$ and
${\Delta_T^{t}}^2_{\min} $, respectively.
Note that the forward peak is narrowed and the backward peak is broadened due to the effect of the boost from the CMS to the LAB frame
which corresponds to the nucleon $N$ at rest in the ${\bar{\rm P}}$ANDA setup.}
\label{FigAngls}
\end{center}
\end{figure}

\section{Pseudoscalar meson production in association with   a high invariant mass lepton pair in $\bar{N} N$  annihilation}

\subsection{Results for $\pi^0$ and $\pi^-$ production }
\label{Ssec-41}

On Figs.~\ref{FigCSpi0} and \ref{FigCSpiminus},
we show our model predictions for the integrated cross section
\be
\frac{d \bar{\sigma} }{dQ^2}({\Delta_T^2}_{\min}) \equiv \int_{u_{\min}}^{u_{\max}} du \int d \theta_{\ell} \frac{d \sigma}{du dQ^2 d \cos \theta_\ell }
\label{Def_itegrated_CS}
\ee
of
$\bar{p}p \rightarrow \ell^+\ell^- \pi^0$
and of
of
$\bar{p}n \rightarrow \ell^+\ell^- \pi^-$
as the function of
$Q^2$,
for several values of
$W^2$ ($W^2=5,\;10$ and $20$ GeV$^2$ ) with the cut  at ${\Delta_T^2}_{\min}=-0.2$ GeV$^2$.
As  phenomenological input, our model for
$\pi N$
TDAs requires the nucleon DAs
$V^p$, $A^p$, $T^p$
at the normalization scale
$\mu^2 \sim Q^2$.
A vast literature exists on the phenomenological solutions for nucleon
DAs (see
{\it e.g.}
\cite{Stefanis_DrNauk,Braun:2006hz}
for the discussion).
In order to quantify the sensitivity of our model prediction on the input nucleon DAs,
we show the cross section estimates for the case of several phenomenological solutions fitting the
nucleon electromagnetic form factor:
Chernyak-Ogloblin-Zhitnitsky (COZ)
\cite{Chernyak:1987nv}%
\footnote{Eq.~(13) of Ref.~\cite{Chernyak:1987nv}.}
(dashed line with long dashes),
King and Sachrajda (KS)
\cite{King:1986wi}%
\footnote{Eq.~(4.6) of Ref.~\cite{King:1986wi}} (solid line),
Braun-Lenz-Wittmann (BLW NLO) model of
\cite{Braun:2006hz}%
\footnote{See Appendix B of Ref.~\cite{Braun:2006hz}.}
(dashed line with medium dashes)
and NNLO modification of BLW model suggested in Ref.~\cite{Lenz:2009ar}%
\footnote{See Table~I of Ref.~\cite{Lenz:2009ar}.} (dashed line with short dashes).
The phenomenological solutions for the nucleon DA presented in the literature
are usually given at a low normalization point $\mu^2=1$ GeV$^2$, while the adequate choice
of the normalization scale within our factorized approach is $\mu^2=Q^2$. In our cross section
estimates presented on Figs.~\ref{FigCSpi0}, \ref{FigCSpiminus} we do not
take systematically into account the evolution effects which are expected to be not very significant
due to the small lever arm in $Q^2$.

To get an estimate of these effects for the kinematics in question
we show on  Fig.~\ref{ScaleD}  the ratio of the integrated cross sections
\be
\frac{d \bar{\sigma} / d Q^2 |_{\mu^2=Q^2}} { d \bar{\sigma}/ d Q^2|_{\mu^2=1 \, {\rm GeV}^2 }}
\ee
for
$\bar{p} p \to \pi^0 \ell^+ \ell^-$
with and without account of the evolution effects for input nucleon DA as the
function of $Q^2$ (with $\alpha_s$ kept fixed). The result is presented for the BLW NNLO input nucleon DA.
For   $Q^2=2 \div 4$ GeV$^2$ the effect of the evolution
results in a mere $40 \%$ decrease of the cross section. This effect is much below the model
dependence due to different phenomenological input.
For other input phenomenological DAs we expect the effect of the evolution to be
of a similar order of magnitude.

Finally   we conclude  that within the considered parameter range,
the cross sections of
$\bar{p} p \to \pi^0 \ell^+ \ell^-$
and
$\bar{p} n \to \pi^- \ell^+ \ell^-$
presented on
Figs.~\ref{FigCSpi0}, \ref{FigCSpiminus}
seem to be large enough for a detailed investigation to be carried
at PANDA@GSI-FAIR.

\begin{figure}[H]
 \begin{center}
 \epsfig{figure=  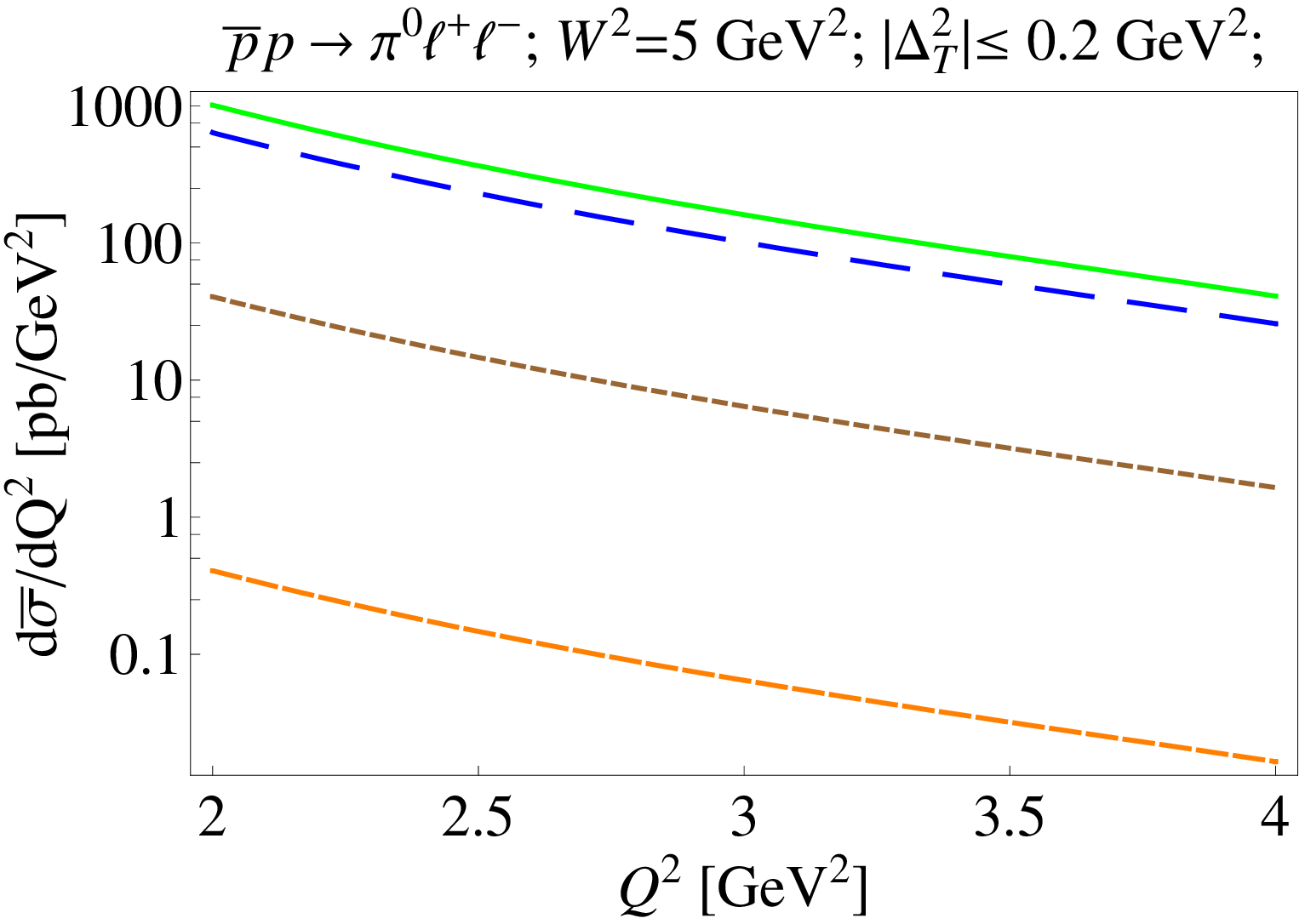 , height=5.3cm}
  \epsfig{figure= 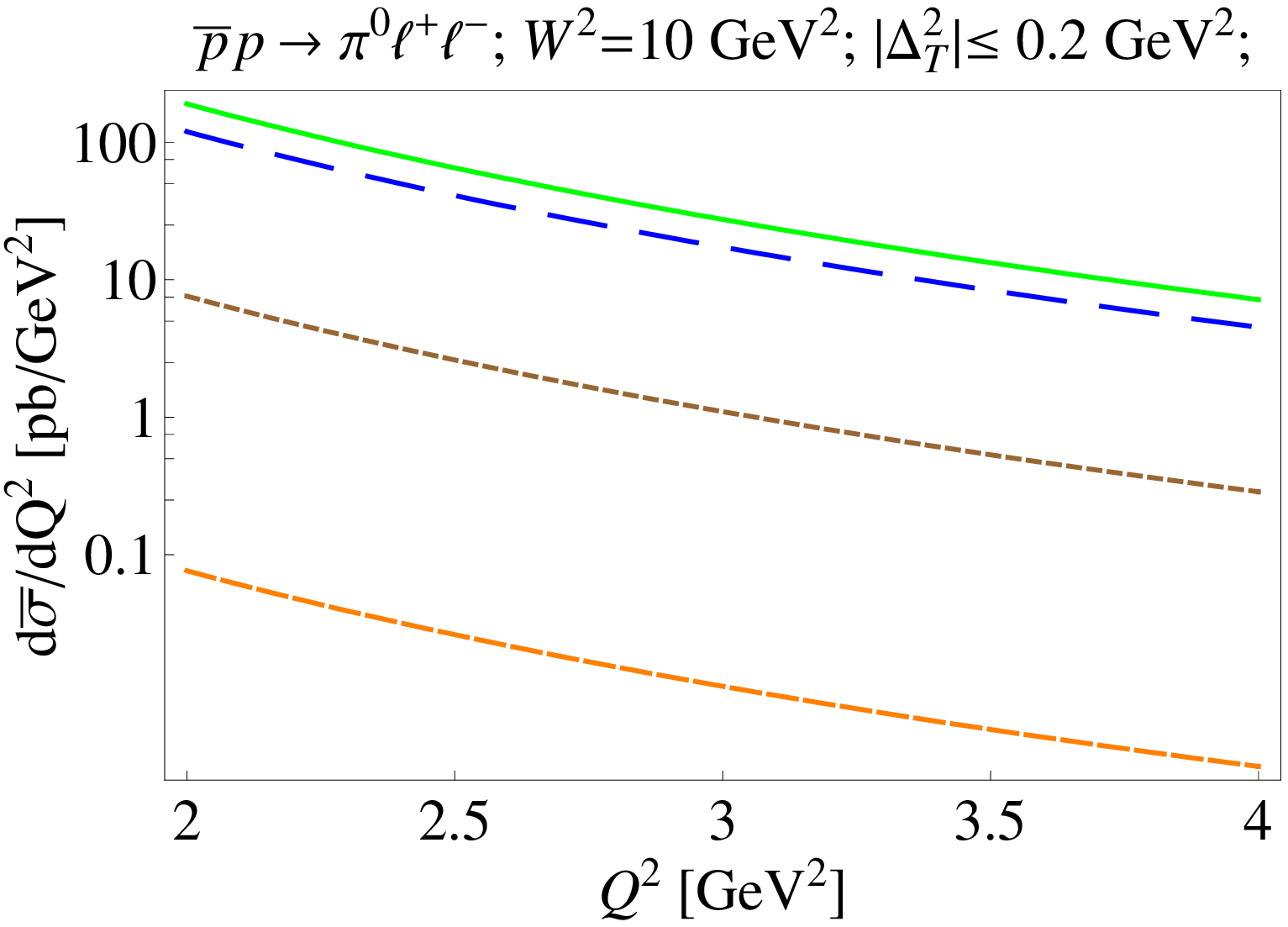 , height=5.3cm}
   \epsfig{figure= 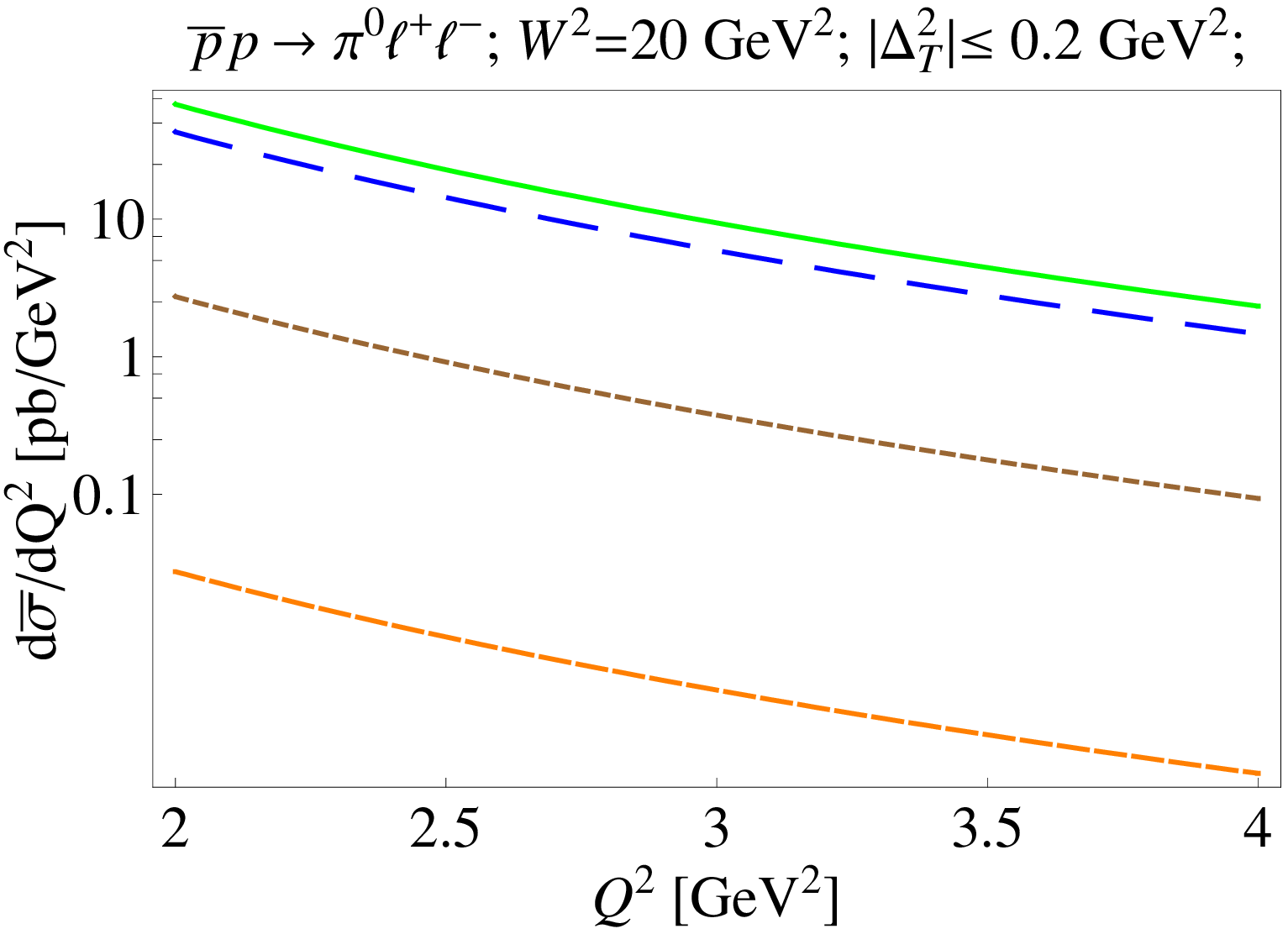 , height=5.3cm}
     \caption{Integrated cross section $d \bar{\sigma}  /dQ^2$ for
     $\bar{p}p \rightarrow \ell^+\ell^- \pi^0$ as a function of $Q^2$ for different values of $W^2=5,\;10$ and $20$ GeV$^2$
for various phenomenological nucleon DA solutions: COZ (long dashes);
     KS (solid line);   BLW NLO
(medium dashes) and NNLO modification~\cite{Lenz:2009ar} of BLW (short dashes).}
\label{FigCSpi0}
\end{center}
\end{figure}

It is worth  mentioning that the cross section that we obtained with  the BLW NLO model input is
much smaller than that in the case of models including next-to-next-to-leading conformal spin
contribution (such as COZ, KS and NNLO modification of BLW). In fact, the
$p\bar{p} \rightarrow \pi^0 \gamma^*$
and
$n\bar{p} \rightarrow \pi^- \gamma^*$
amplitudes turn  to be zero at the leading twist accuracy once we employ the asymptotic form of the nucleon DA
as input. This zero is reminiscent of the zero for the proton electromagnetic form factor with the asymptotic DA
\cite{Lepage:1980}.
The cross section thus turns out to be small for the BLW NLO as well as for the Bolz-Kroll (BK)
\cite{Bolz:1996sw}
input nucleon DAs which are known to be rather close to the asymptotic form of the nucleon DA.

\begin{figure}[H]
 \begin{center}
 \epsfig{figure=  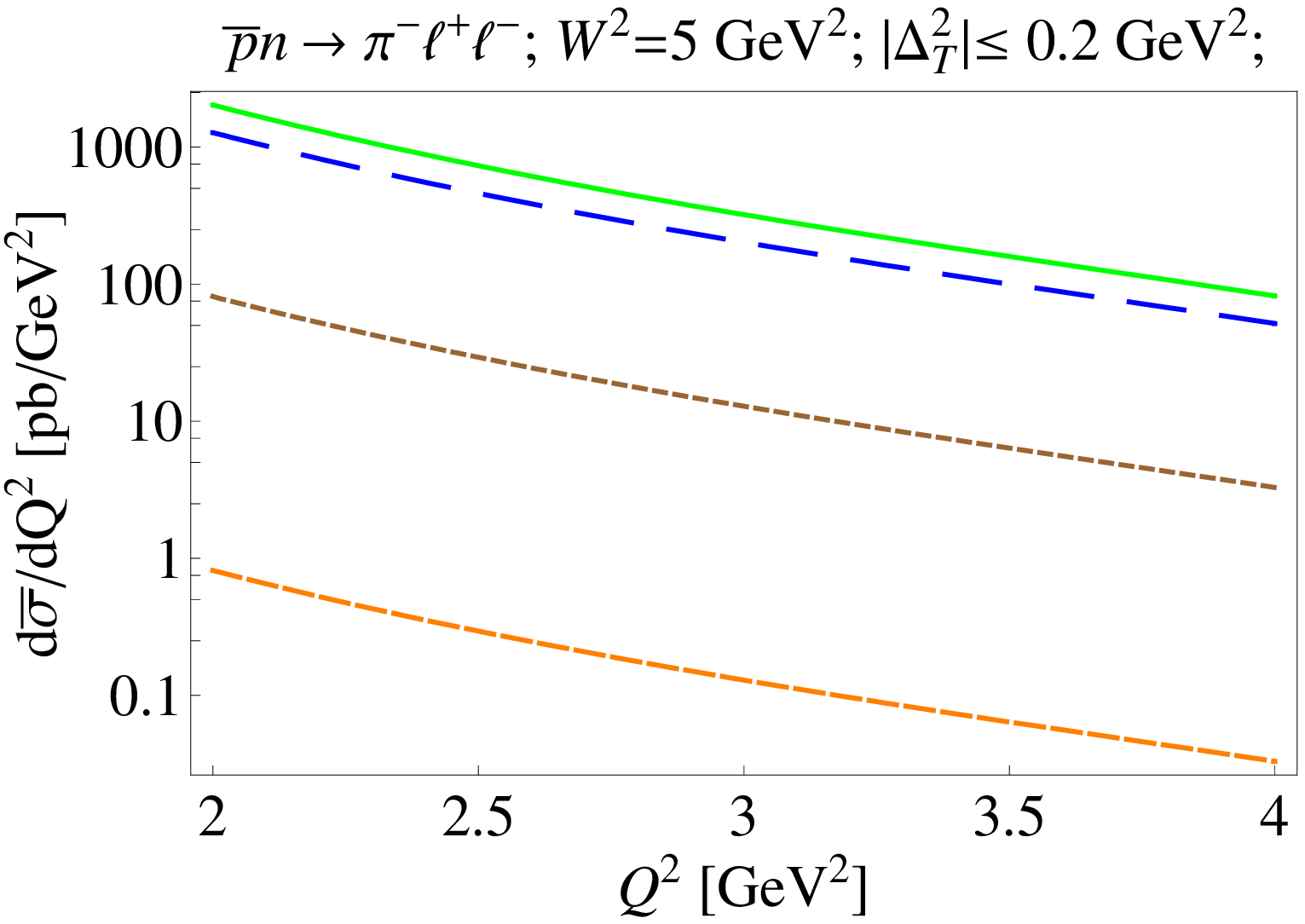 , height=5.3cm}
  \epsfig{figure= 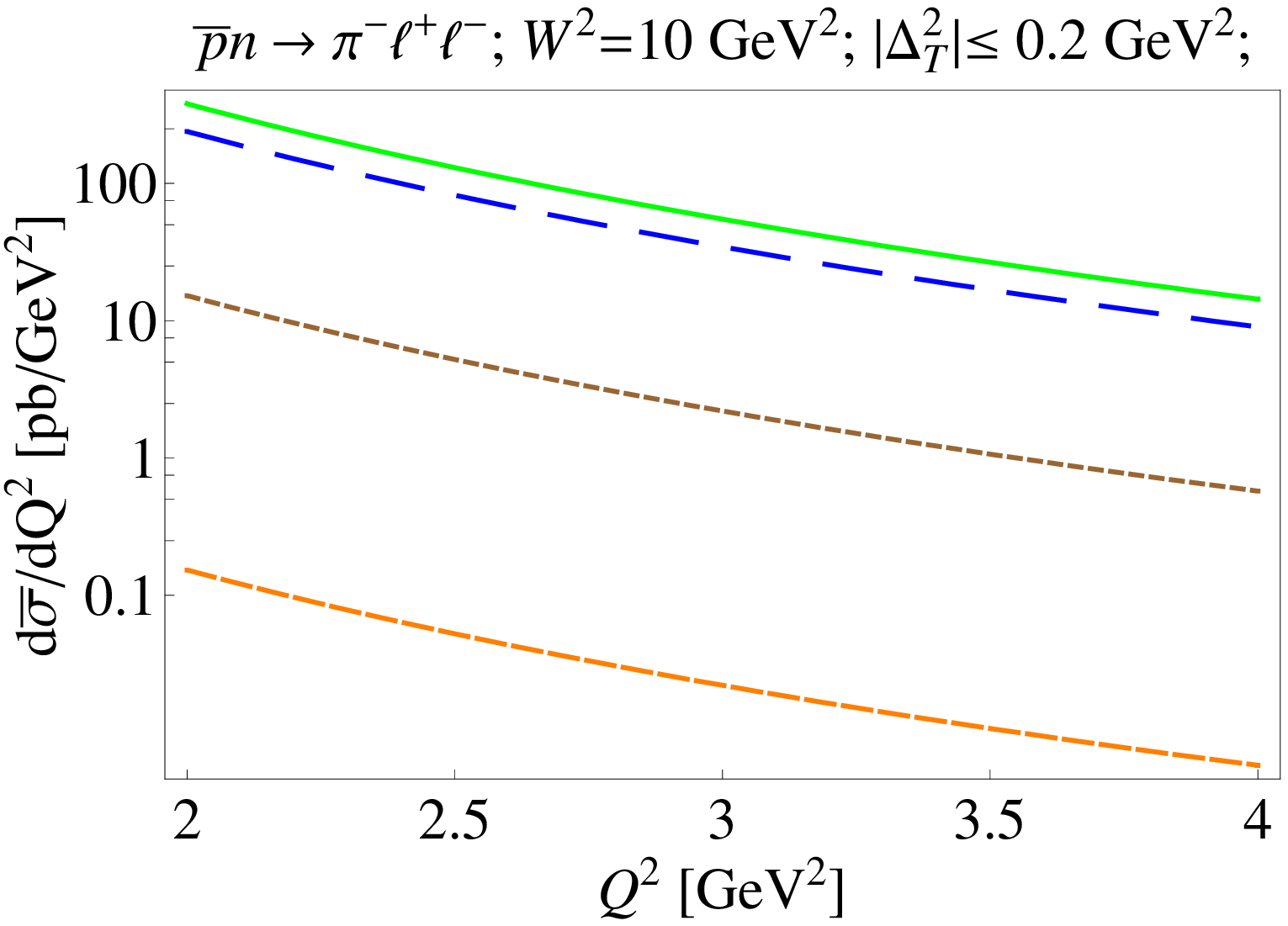 , height=5.3cm}
   \epsfig{figure= 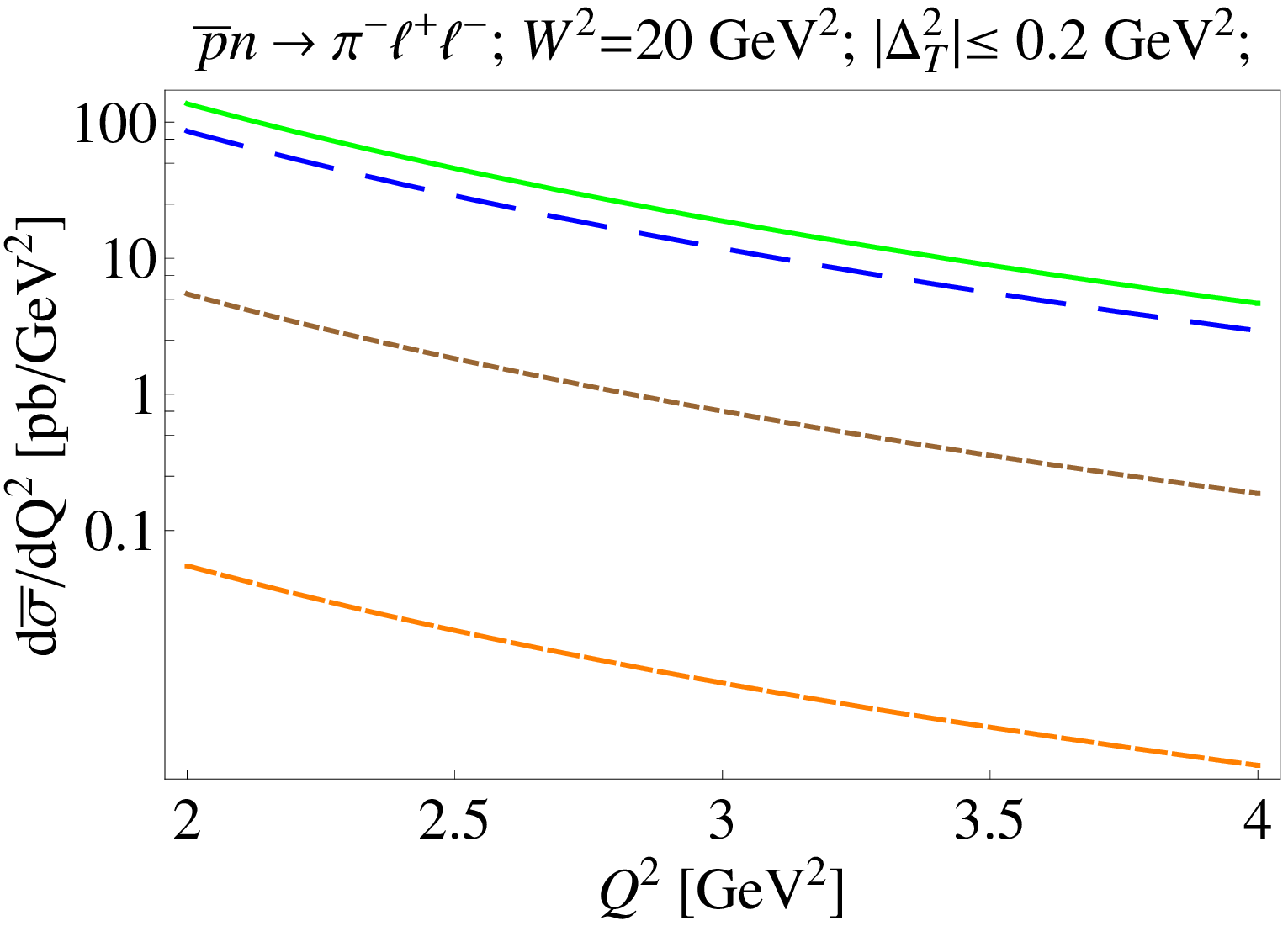 , height=5.3cm}
     \caption{Integrated cross section $d \bar{\sigma} /dQ^2$ for
     $\bar{p}n \rightarrow \ell^+\ell^- \pi^-$ as a function of $Q^2$, for different values of $W^2=5,\;10$ and $20$ GeV$^2$ for various
 phenomenological nucleon DA solutions: COZ (long dashes);
     KS (solid line);   BLW NLO
(medium dashes) and   NNLO modification~\cite{Lenz:2009ar} of BLW   (short dashes). }
\label{FigCSpiminus}
\end{center}
\end{figure}

\begin{figure}[H]
 \begin{center}
 \epsfig{figure=  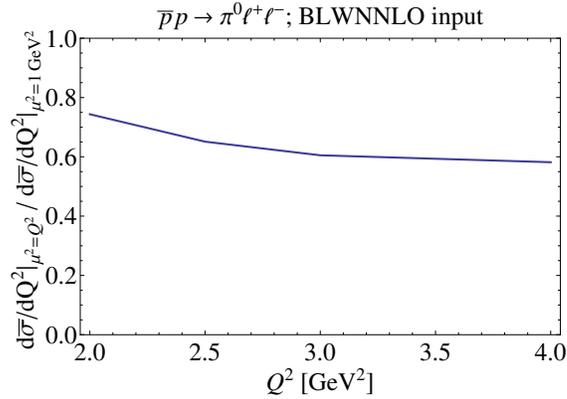 , height=5.3cm}
      \caption{ The ratio of the
integrated cross sections
$
\frac{d \bar{\sigma} / d Q^2 |_{\mu^2=Q^2}} { d \bar{\sigma}/ d Q^2|_{\mu^2=1 \, {\rm GeV}^2 }}
$
with and without taking account of the evolution effects as a
function of $Q^2$.
}
\label{ScaleD}
\end{center}
\end{figure}

In Fig.~\ref{Fig_peaks}, we show
the differential cross section $d \sigma /dQ^2 d \cos \theta_\pi^*$ for
$\bar{p}p \rightarrow \ell^+\ell^- \pi^0$ as a function of $\cos \theta_\pi^*$,
for $W^2=5$ GeV$^2$ and  $Q^2=2.5$ GeV$^2$, both for the near forward and backward
factorization regimes.
As a consequence of $C$-invariance, the cross section of
(\ref{BarNNannihilation reaction}) within the $t$-channel factorization mechanism
can be obtained from that within the $u$-channel factorization mechanism
with the  change (\ref{tu_change}) (see discussion in App.~\ref{App_B3}).
The COZ solution for the nucleon DAs is used here as  numerical input.
The forward and backward peaks produced within the suggested factorization mechanism are clearly visible.
They look perfectly symmetric in the CMS frame. However, it is worth  mentioning that
when boosting from the CMS to the LAB frame (which corresponds to the nucleon $N$ at rest in the ${\bar{\rm P}}$ANDA setup),
the forward peak is narrowed and the backward peak is broadened (see Fig.~\ref{FigAngls}).

\begin{figure}[H]
 \begin{center}
 \epsfig{figure=  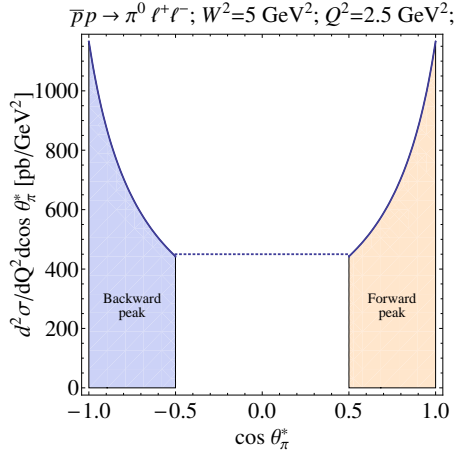 , height=6.0cm}
  \caption{Differential cross section $d \sigma /dQ^2 d \cos \theta_\pi^*$ for
     $\bar{p}p \rightarrow \ell^+\ell^- \pi^0$ as a function of $\cos \theta_\pi^*$
     for $W^2=5$GeV$^2$ and  $Q^2=2.5$GeV$^2$. Forward and backward peaks are clearly visible.
     COZ solution for the nucleon DAs is used as the numerical input. Dotted region denotes scattering
     over large angles in which the present factorization description does not apply.}
\label{Fig_peaks}
\end{center}
\end{figure}

Obviously, the $\pi N$ TDAs in the nucleon pole model have purely ERBL support and
turn into zero at the borders of the ERBL domain.
In a general model in which $\pi N$ TDAs do not vanish at the borders of the ERBL domain
(so-called crossover trajectories $w=-\xi$, $v= \pm \xi'$),
important contributions into the cross section may come from the convolutions
with coefficient functions that are highly singular at the crossover trajectories.

To estimate this effect we consider the two component model for $\pi N$ TDAs  \cite{Lansberg:2011aa}
which includes the spectral part (formulated in terms of quadruple distributions)
and the nucleon pole part, which is the analogue of the $D$-term familiar from the GPD case.
The spectral part is nonzero in both  the ERBL and the Dokshitzer-Gribov-Lipatov-Altarelli-Parisi (DGLAP)-like domains and does not vanish at the cross
over trajectories. This model is an attempt to maximally take into account the general requirements
(following from the underlaying fundamental field theory), such as the following:
\begin{itemize}
\item polynomiality property and the support property which are the consequences of the Lorentz invariance;
\item permutation symmetry and the isotopic invariance;
\item chiral properties manifest through the soft pion theorem.
\end{itemize}
Nevertheless the two component model of Ref.~\cite{Lansberg:2011aa} is still very flexible.
In particular, the factorized Ansatz for the quadruple distributions and
the shape of the profile function are ad hoc assumptions. So we
can rather make a model dependent estimate of the contributions of the cross over trajectories and of the DGLAP region   
into the cross section.
As the nucleon pole contribution dies out for $\xi=1$ ,the spectral part contribution
becomes relatively more important for higher $\xi$.

In Fig.~\ref{FigCS_SP} we show the effect of adding the contribution of the spectral part
for the integrated cross section $d \bar{\sigma} / d Q^2$ of $p \bar{p} \to \pi^0 \ell^+ \ell^-$
presented inn Fig.~\ref{FigCSpi0}.
Thin dashes show the pure nucleon pole part contribution; thick dashed lines
show the sum (at the amplitude level) of two contributions. The COZ phenomenological
solution for the nucleon DA is used as the numerical input. For different phenomenological
inputs the effect of adding the spectral part is similar.

\begin{figure}[H]
 \begin{center}
 \epsfig{figure=  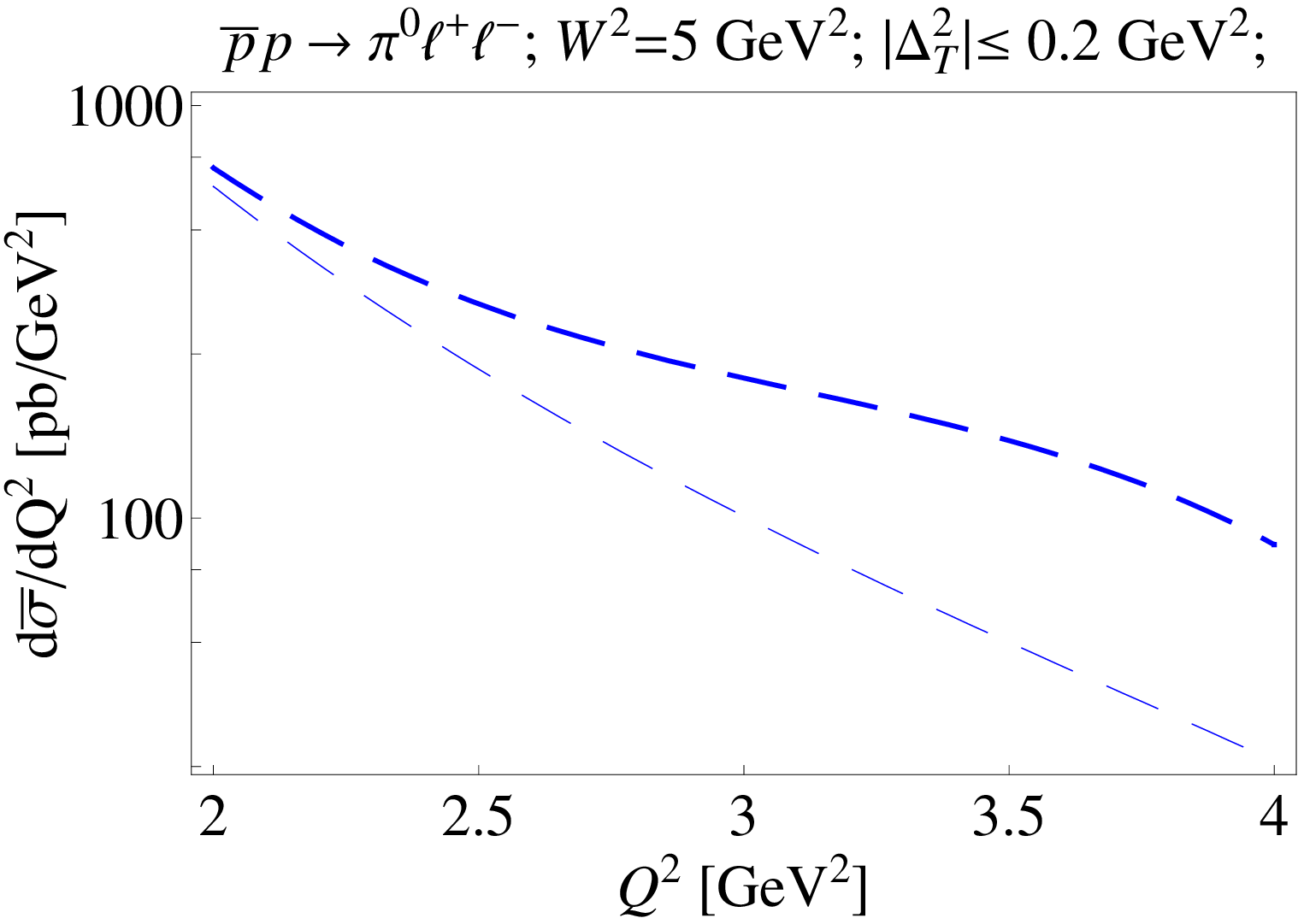 , height=5.3cm}
  \epsfig{figure= 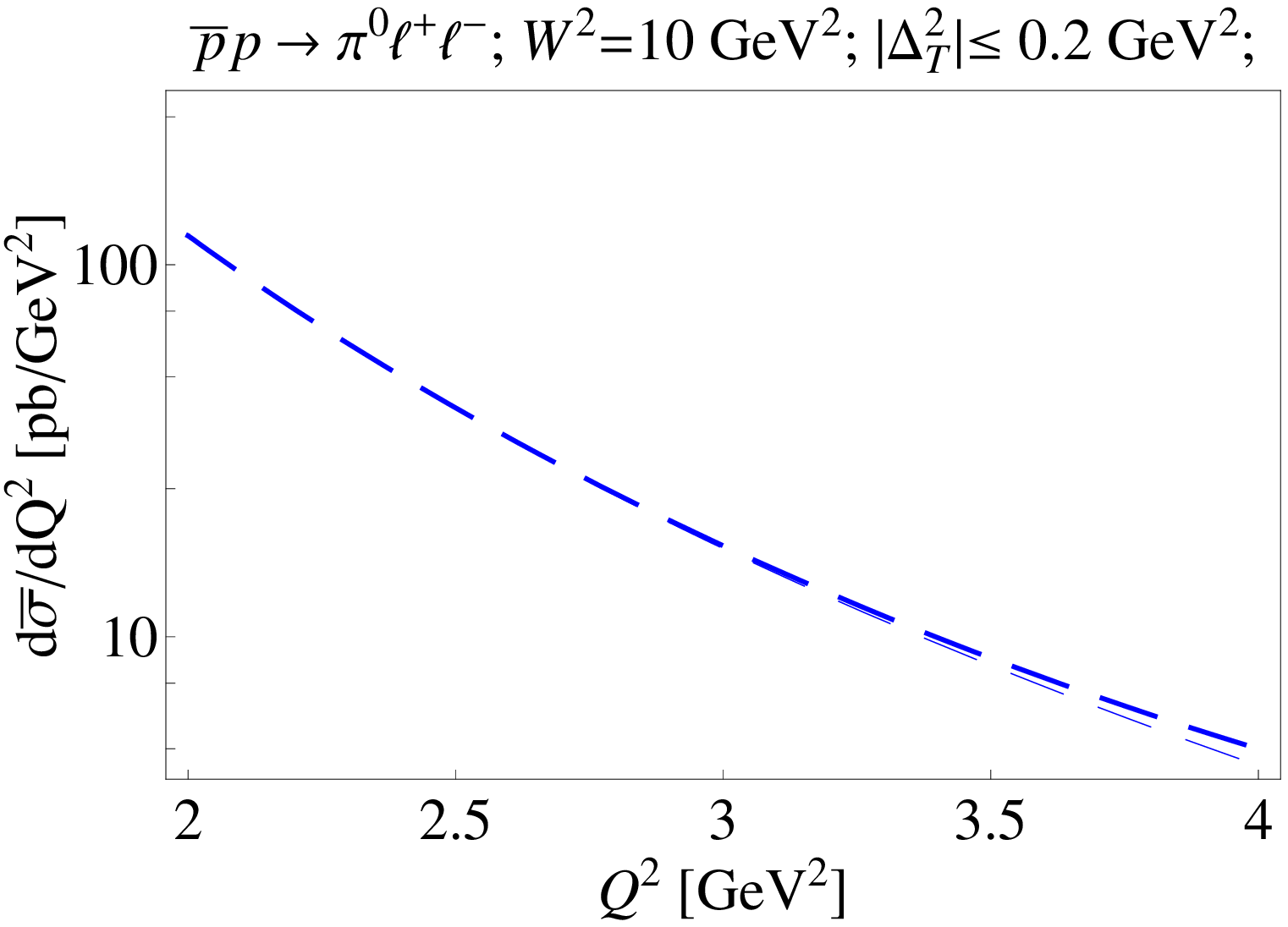 , height=5.3cm}
      \caption{The integrated cross section $d \bar{\sigma}  /dQ^2$ of
     $\bar{p}p \rightarrow \ell^+\ell^- \pi^0$ as a function of $Q^2$ for   $W^2=5$ GeV$^2$  (left panel) and $W^2=10$ GeV$^2$ (right panel)
in the two component model of $\pi N$ TDAs  (thick dashed line) v.s. the nucleon pole model (thin dashed line).
COZ nucleon DA is used as the numerical input.}
\label{FigCS_SP}
\end{center}
\end{figure}

One may see that for
$W^2=5$ GeV$^2$,
adding the contribution of the spectral part   results, at most, in a  factor of $2$ increase of the
cross section for
$Q^2 \sim 4$ GeV$^2$
(this kinematical region corresponds to rather large values of
$\xi \sim 0.6$).
This effect is not very important in view of the large uncertainty
due to different input phenomenological DAs. Therefore the account of the spectral part contribution
together with the question of appropriate modelling of quadruple distributions will become a topical question
after a stable signal from the backward (and forward) regimes will be available from PANDA.
Moreover, for $W^2=10$ GeV$^2$ the $Q^2$ interval in question corresponds to rather small
values of $\xi \sim 0.1$, for which  the nucleon pole part totally dominates over the spectral part within the present
model.

\subsection{Results for $\eta $ production
}

In an analogous way, we may consider near backward (and forward) $\eta$ meson production in association with a lepton pair
\be
N(p_p)+\bar{N}(p_{\bar{p}}) \rightarrow \gamma^*(q)+\eta(p_\eta)  \rightarrow \ell^+(p_{\ell^+})+\ell^-(p_{\ell^-})+\eta(p_\eta).
\label{React_PANDA_eta}
\ee
The nucleon pole model for $\eta N$ TDAs is similar to that for $\pi N$ TDAs (\ref{Nucleon_exchange_contr_VAT})  with the obvious
change of phenomenological coupling $g_{\pi NN} \rightarrow g_{\eta NN}$.
Estimates of  $g_{\pi NN}$ and
$g_{\eta NN}$
phenomenological couplings  taken from
Table~9.1 of
Ref.~\cite{Dumbrajs:1983jd}
give
$\frac{g_{\eta NN}^2}{g_{\pi NN}^2} \sim 0.3$.
The formulas for the
$\bar{p} p \to \gamma^* \eta \to \ell^+ \ell^- \eta$
are obtained from the relevant formulas of Sec.~\ref{Sec-3} with the obvious
change of values of the masses and the couplings.
In Fig.~\ref{FigCSeta}, we show the results for the integrated cross section
$d \sigma^{\rm int} /dQ^2$
for
$\bar{p}p \rightarrow \ell^+\ell^-\eta$
as a function of
$Q^2$,
for different
$W^2=5,\;10$ and $20$ GeV$^2$
for various phenomenological nucleon DA solutions:
COZ (long dashes), KS (solid line),  BLW NLO
(medium dashes)   and  NNLO modification~\cite{Lenz:2009ar}
of BLW model (short dashes).

\begin{figure}[H]
 \begin{center}
 \epsfig{figure=  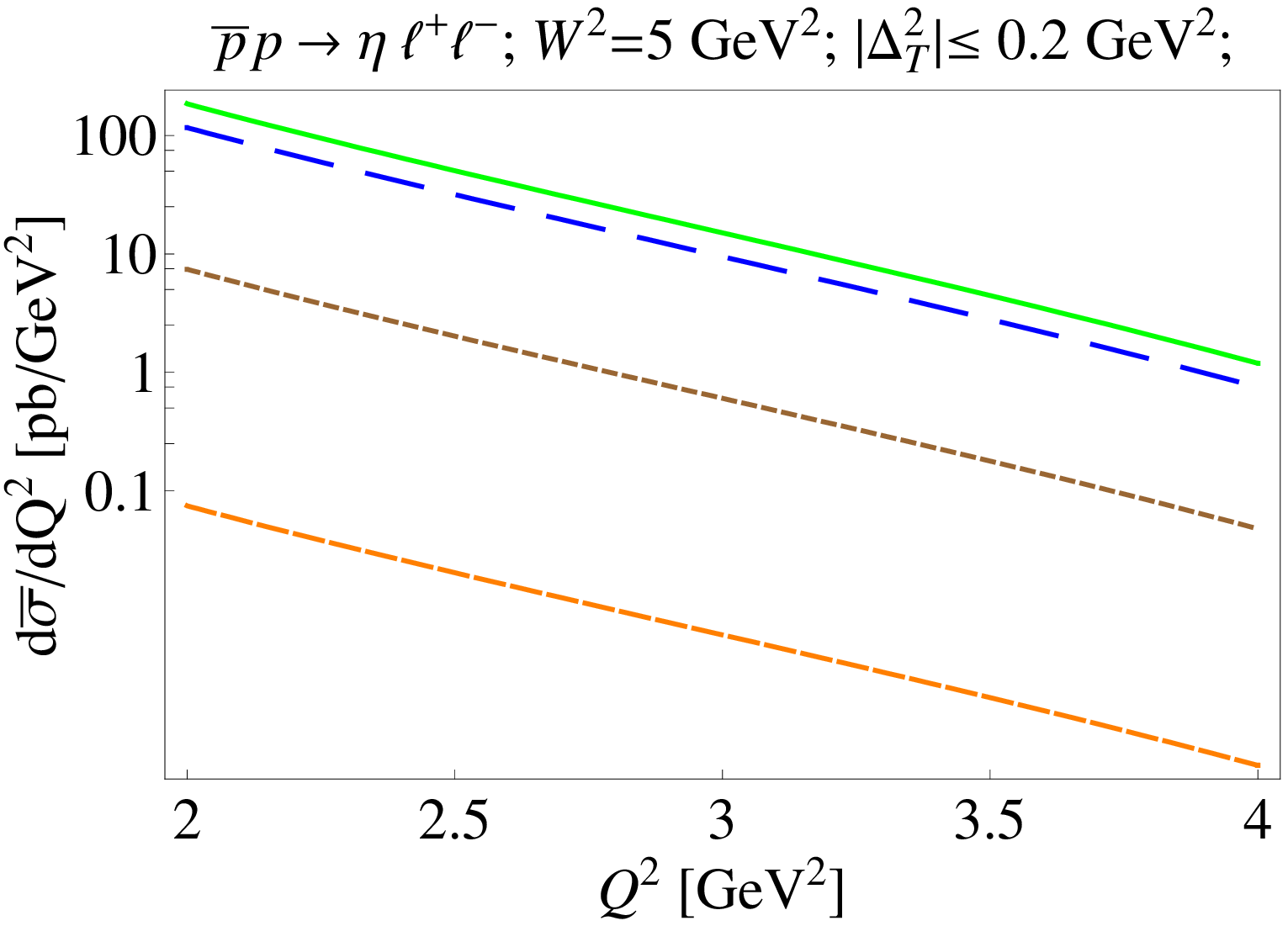 , height=5.3cm}
  \epsfig{figure= 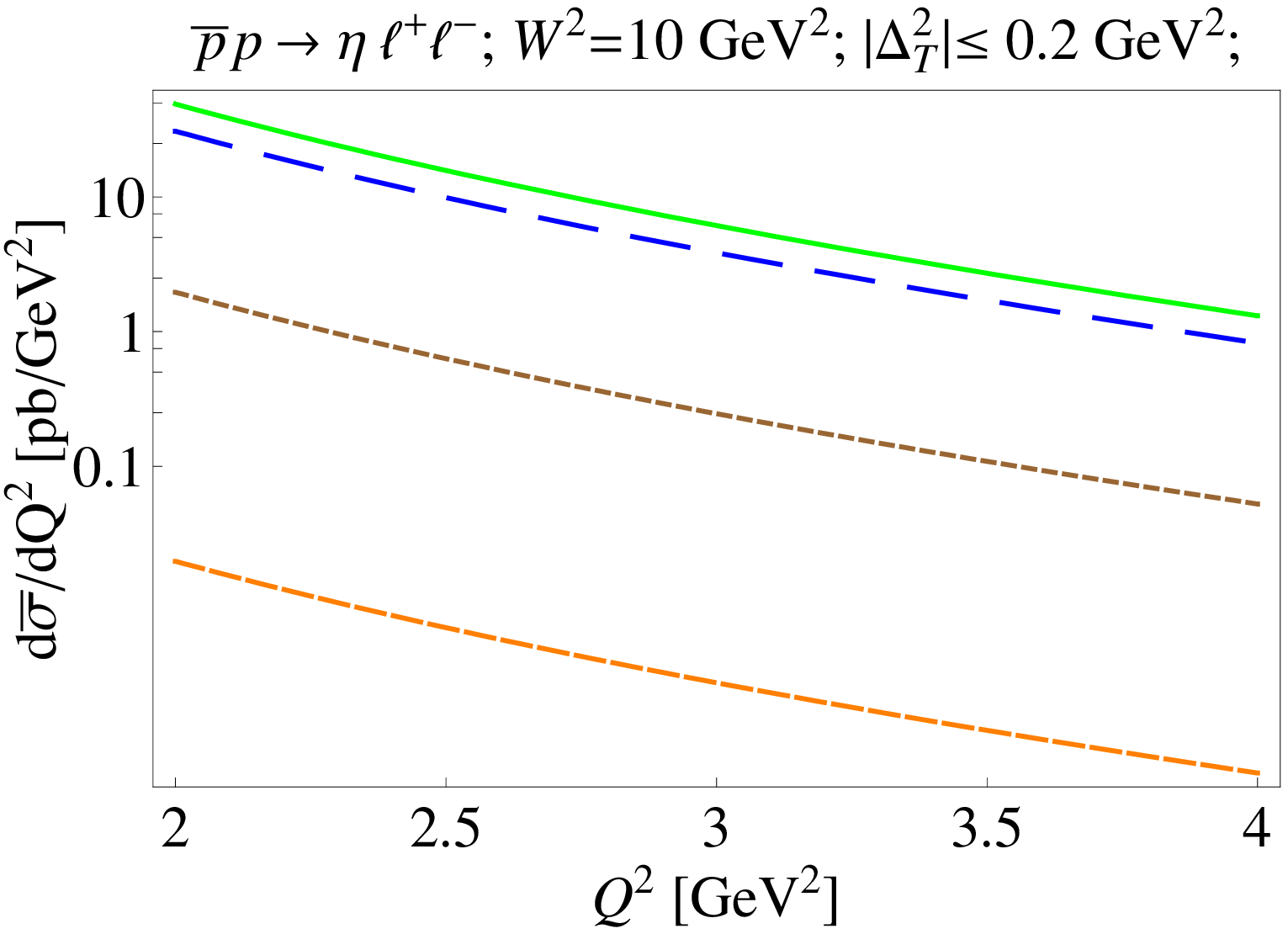 , height=5.3cm}
   \epsfig{figure= 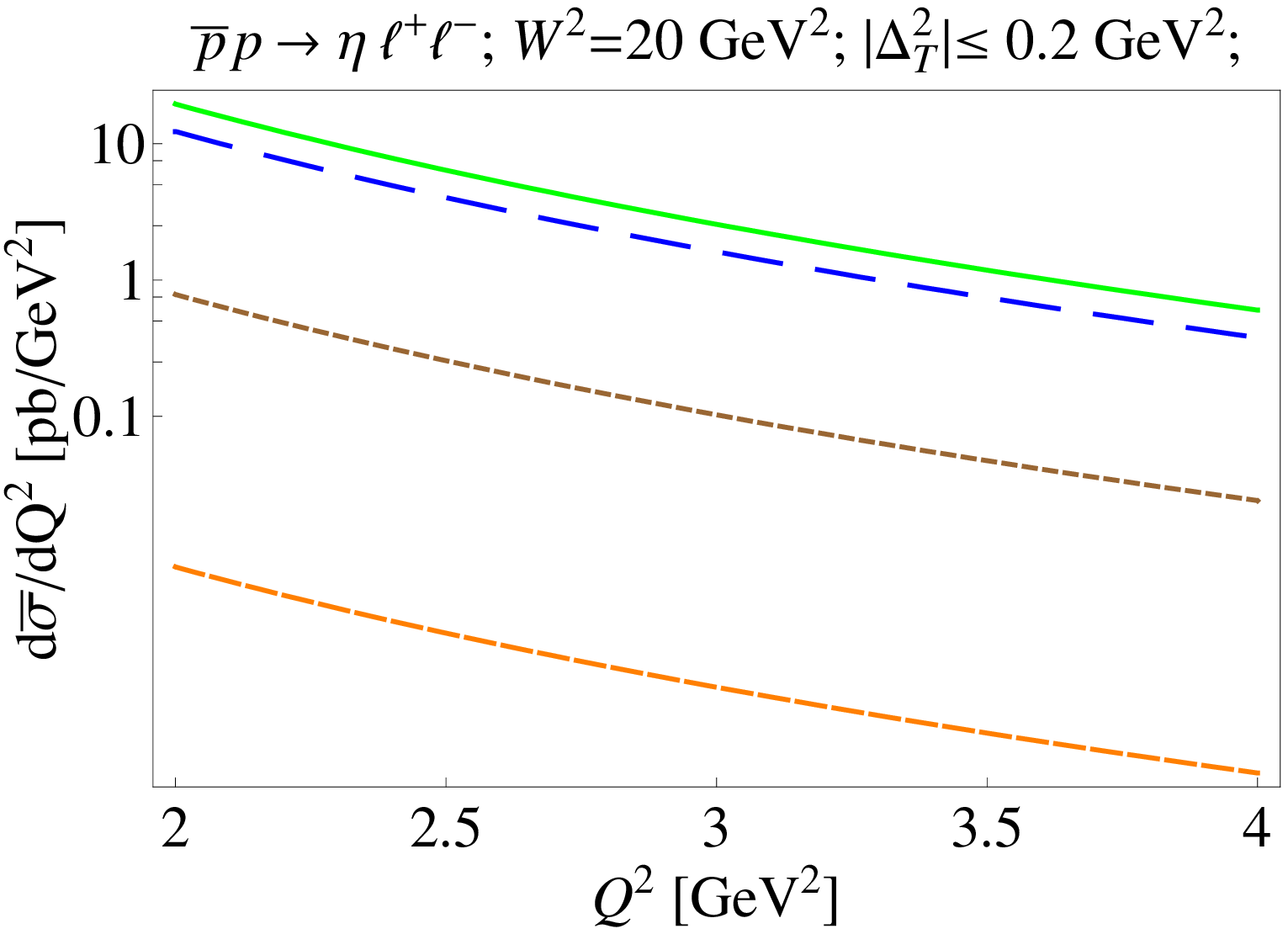 , height=5.3cm}
     \caption{Integrated cross section
     $d \bar{\sigma}  /dQ^2$ for
     $\bar{p}p \rightarrow \ell^+\ell^- \eta$
     as a function of $Q^2$ for different values of
     $W^2=5,\;10$ and $20$ GeV$^2$
     for various phenomenological nucleon DA solutions: COZ (long dashes);
     KS (solid line);   BLW NLO
(medium dashes) and  NNLO modification~\cite{Lenz:2009ar} of BLW  (short dashes). }
\label{FigCSeta}
\end{center}
\end{figure}

\section{Conclusions}
\label{Sec-Concl}

In this work, we have estimated the cross section for pseudoscalar meson production
in association with a high invariant mass lepton pair in nucleon antinucleon annihilation.
We have performed this analysis in the
framework of the QCD collinear factorization. We used the model for
the TDAs which was introduced recently in \cite{Pire:2011xv} and for which the major input is the parametrization
of the  nucleon distribution
amplitudes. For most of the choices, the magnitudes of the cross sections are sufficient to be
measurable with the luminosity foreseen at the ${\bar{\rm P}}$ANDA experiment
at GSI-FAIR.  On the other hand, it turns out that the cross section depends much on
the parametrization used for these DAs, which may be analyzed in a positive way as
an indication that the experimental study of these processes will help to distinguish
between the different models for these DAs. Note that these different models all
claim to be in reasonable agreement with the nucleon form factor measurements.
We are, however, still far from  proposing a strategy for a
model-independent extraction of the TDAs from experimental data. More observables,
and in particular spin asymmetries, should be discussed. Coefficient functions
should  also be calculated at next-to-leading order and the effect of the $Q^2$ evolution
of the TDAs should also be studied in details.  Note that these NLO order effects are known to
be different in timelike and spacelike processes
\cite{Muller:2012yq}.
Meanwhile, the preparation for the experimental  analysis requires detailed background studies (such
as that of the  final states where a
$\pi^+ \pi^-$
meson pair mimics the lepton pair) and a careful estimation of detector efficiency.
Although  we concentrated here on pseudoscalar ($\pi$ and $\eta$) meson production, this
framework can be extended to the production of vector mesons ($\rho$, $\omega$) which will uncover
complementary aspects of hadronic structure both experimentally and theoretically.
Also a purely
hadronic description of $\bar{p} p \to \pi  \ell^+ \ell^-$  is proposed in
Refs. \cite{Adamuscin:2007iv} and \cite{Guttmann:2012sq}, respectively, in a Born model
and a Regge model description of baryonic exchanges.

\section*{Acknowledgements}
We are   thankful to Thierry Hennino, Frank Maas and Manuel
Zambrana for constructive and motivating discussions.
This work is partly supported by the Polish Grant NCN
No. DEC-2011/01/B/ST2/03915, the French-Polish collaboration agreement
Polonium, the
P2IO consortium and the Joint Research Activity ”Study of Strongly
Interacting Matter”
(acronym HadronPhysics3, Grant Agreement n.283286) under the Seventh
Framework Programme of the European Community and by the COPIN-IN2P3 Agreement.

\setcounter{section}{0}
\setcounter{equation}{0}
\renewcommand{\thesection}{\Alph{section}}
\renewcommand{\theequation}{\thesection\arabic{equation}}

\section{Kinematics}
\label{App_A}
Below we review the relevant kinematical quantities discussed in
\cite{Lansberg:2007se}
for the two possible factorization regimes for the reaction (\ref{BarNNannihilation reaction})
involving nucleon to meson and antinucleon to meson TDAs.

\subsection*{$u$-channel factorization regime}
Within the conventions natural for the ${\bar{\rm P}}$ANDA  setup,
the $u$-channel factorization regime (see the left panel of Fig.~\ref{Fig1}) in which
$|u|$ is small corresponds to the meson
moving in the direction of the initial proton  (backward direction).

 The $z$-axis is chosen along the colliding nucleon-antinucleon.
We define the light-cone vectors
$p^u$
and
$n^u$
such that
$2 p^u \cdot n^u =1$.
We introduce the average momentum and $u$-channel momentum transfer
\be
P^u=\frac{1}{2}(p_{\mathcal{M}}+p_N); \ \ \ \Delta^u=p_{\mathcal{M}}-p_N; \ \ \ {\Delta^u}^2 \equiv u.
\ee
${\Delta^u}_T$ is the transverse component of
$\Delta^u$ ($ {\Delta^u_T}^2 \le 0$). We define the skewness
variable
\be
\xi^u=- \frac{\Delta^u \cdot n^u}{2 P^u \cdot n^u}.
\label{def_xiu}
\ee

The following Sudakov decomposition is valid for the momenta of the reaction (\ref{React_PANDA}):
\be
&&
p_N= (1+\xi^u) p^u + \frac{M^2}{1+\xi^u} n^u; \nonumber \\ &&
p_{\bar{N}}= \frac{2M^2(1+\xi^u)}{W^2-2 M^2+W \sqrt{W^2-4M^2}}p^u+ \frac{W^2-2 M^2+W \sqrt{W^2-4M^2}}{2(1+\xi^u)}n^u;
\nonumber \\ &&
p_{\mathcal{M}}=(1-\xi^u)p^u+ \frac{m^2-{\Delta^u_T}^2}{1-\xi^u}n^u+{\Delta^u_T};
\nonumber \\ &&
\Delta^u=-2 \xi^u p^u+ \left[ \frac{m^2-{\Delta^u_T}^2}{1-\xi^u}- \frac{M^2}{1+\xi^u} \right] n^u+ {\Delta^u_T} ;
\nonumber \\ &&
q = \left[ 2 \xi^u + \frac{M^2}{W^2} (1+\xi^u) + O(1/W^4) \right] p^u+
\left[
\frac{W^2-M^2}{1+\xi^u} -\frac{m^2-{\Delta^u_T}^2}{1-\xi^u} + O(1/W^2)
\right] n^u-\Delta^u_T, \nonumber \\ &&
\ee
where $M$ ($m$) denotes nucleon (meson) mass.

In the nucleon rest frame (which corresponds to the ${\bar{\rm P}}$ANDA laboratory frame)
the light-cone vectors $p^u$
and
$n^u$
read
\be
p^u|_{N \; {\rm rest}}=\frac{M}{2(1+\xi^u)} \{1,0,0,-1\}; \ \ \ n^u|_{N \; {\rm rest}}=\frac{1+\xi^u}{2M} \{1,0,0,1\}.
\ee
With the help of the appropriate boost
we establish the expressions for the light-cone vectors
in the
$N \bar{N}$
CMS
\be
p^u|_{\bar{N}N \; {\rm CMS}}=\{\alpha^u,0,0,-\alpha^u \}; \ \ \ n^u|_{\bar{N}N \; {\rm CMS}}=\{\beta^u,0,0, \beta^u \},
\ee
where
\be
\alpha^u=\frac{W+\sqrt{W^2-4 M^2}}{4 (1+\xi^u)}; \ \ \ \beta^u= \frac{\left(W-\sqrt{W^2-4 M^2} \right) (1+\xi^u)}{4 M^2}.
\label{Def_alpha_beta_t}
\ee

The meson $\mathcal{M}$ scattering angle in the $N \bar{N}$ CMS for the $u$-channel factorization regime then can be expressed as:
\be
\cos \theta_{\mathcal{M}}^*=\frac{-(1-\xi^u)\alpha^u+\frac{m^2-{\Delta^u_T}^2}{1-\xi^u} \beta^u}
{\sqrt{(-(1-\xi^u)\alpha^u+\frac{m^2-{\Delta^u_T}^2}{1-\xi^u} \beta^u)^2-{\Delta^u_T}^2}}.
\label{CosThetaPi_tregime}
\ee
One may check that for ${\Delta^u_T}^2=0$ indeed $\cos \theta_\mathcal{M}^*=-1$, which means backward
scattering.

We also quote some useful relations for the kinematical quantities:
\be
&&
{\Delta^u_T}^2= \frac{1-\xi^u}{1+\xi^u} \left(u- 2\xi^u \left[\frac{M^2}{1+\xi^u} -\frac{m^2}{1-\xi^u} \right] \right);
\nonumber  \\ &&
Q^2 \equiv q^2 =\frac{2 \xi^u}{1+\xi^u} W^2+u-3M^2 +\frac{4M^2}{1+\xi^u}+O(1/W^2);
\nonumber  \\ &&
t=(p_{\mathcal{M}}-p_{\bar{N}})^2=- \frac{1-\xi^u}{1+\xi^u} W^2+m^2-M^2+\frac{4M^2}{1+\xi^u}+ O(1/W^2).
\label{Kin_quant_t}
\ee
This allows to express $\xi^u$ as
\be
\xi^u \simeq \frac{Q^2-u-M^2}{2W^2-Q^2+u-3 M^2}.
\ee

\subsection*{$t$-channel factorization regime}
The $t$-channel factorization regime (see the right panel of Fig.~\ref{Fig1}) in which
$|t|$ is small corresponds to the meson
moving in the direction of the initial antiproton (forward direction).

We define the light-cone vectors
$p^t$
and
 $n^t$
such that
$2 p^t \cdot n^t =1$.
\be
P^t=\frac{1}{2}(p_{\mathcal{M}}+p_{\bar{N}}); \ \ \ \Delta^t=p_{\mathcal{M}}-p_{\bar{N}}; \ \ \
{\Delta^t}^2 \equiv t.
\ee
${\Delta^t}_T$ is the transverse component of
$\Delta^t$ ($\Delta^t_T \cdot \Delta^t_T ={\Delta^t_T}^2 \le 0$). We define the skewness
variable
\be
\xi^t=- \frac{\Delta^t \cdot n^t}{2 P^t \cdot n^t}.
\label{defxit}
\ee

The following Sudakov decomposition is valid for the momenta of particles:
\be
&&
p_{\bar{N}}= (1+\xi^t) p^t + \frac{M^2}{1+\xi^t} n^t; \nonumber \\ &&
p_{N}= \frac{2M^2(1+\xi^t)}{W^2-2 M^2+W \sqrt{W^2-4M^2}}p^t+ \frac{W^2-2 M^2+W \sqrt{W^2-4M^2}}{2(1+\xi^t)}n^t;
\nonumber \\ &&
p_{\mathcal{M}}=(1-\xi^t)p^t+ \frac{m^2-{\Delta^t_T}^2}{1-\xi^t}n^t+{\Delta^t_T};
\nonumber \\ &&
\Delta^t=-2 \xi^t p^t+ \left[ \frac{m^2-{\Delta^t_T}^2}{1-\xi^t}- \frac{M^2}{1+\xi^t} \right] n^t+ {\Delta^t_T} ;
\nonumber \\ &&
q = \left[ 2 \xi^t + \frac{M^2}{W^2} (1+\xi^t) + O(1/W^4) \right] p^t+
\left[
\frac{W^2-M^2}{1+\xi^t} -\frac{m^2-{\Delta^t_T}^2}{1-\xi^t} + O(1/W^2)
\right] n^t-\Delta^t_T. \nonumber \\ &&
\ee

In the antinucleon rest frame
the light-cone vectors $p^t$
and
$n^t$
are
\be
p^t|_{\bar{N} \; {\rm rest}}=\frac{M}{2(1+\xi^t)} \{1,0,0,1\}; \ \ \ n^t|_{\bar{N} \; {\rm rest}}=\frac{1+\xi^t}{2M} \{1,0,0,-1\}.
\ee
The explicit expressions for the light-cone vectors $p^t$ and $n^t$ in the $\bar{N}N$ CMS read:
\be
p^t|_{\bar{N}N \; {\rm CMS}}=\{\alpha^t,0,0, \alpha^t \}; \ \ \ n^t|_{\bar{N}N \; {\rm CMS}}=\{\beta^t,0,0, -\beta^t \},
\ee
where $\alpha^t$ and $\beta^t$ are defined  as
\be
\alpha^t=\frac{W+\sqrt{W^2-4 M^2}}{4 (1+\xi^t)}; \ \ \ \beta^t= \frac{\left(W-\sqrt{W^2-4 M^2} \right) (1+\xi^t)}{4 M^2}.
\label{Def_alpha_beta_u}
\ee

The meson $\mathcal{M}$ scattering angle in the $N \bar{N}$ CMS for the $t$-channel factorization regime then  expresses as:
\be
\cos \theta_\mathcal{M}^*=\frac{ (1-\xi^t)\alpha^t-\frac{m^2-{\Delta^t_T}^2}{1-\xi^t} \beta^t}
{\sqrt{((1-\xi^t)\alpha^t-\frac{m^2-{\Delta^t_T}^2}{1-\xi^t} \beta^t)^2-{\Delta^t_T}^2}}.
\label{Cos_theta_u}
\ee
Again one may check that for ${\Delta^t_T}^2=0$, indeed, $\cos \theta_\mathcal{M}^*=1$,  which means forward
scattering.

 We also work out the relations:
\be
&&
{\Delta^t_T}^2= \frac{1-\xi^t}{1+\xi^t} \left(t- 2\xi^t \left[\frac{M^2}{1+\xi^t} -\frac{m^2}{1-\xi^t} \right] \right);
\nonumber  \\ &&
Q^2 \equiv q^2 =\frac{2 \xi^t}{1+\xi^t} W^2+t-3M^2 +\frac{4M^2}{1+\xi^t}+O(1/W^2);
\nonumber  \\ &&
u=(p_{\mathcal{M}}-p_{N})^2=- \frac{1-\xi^t}{1+\xi^t} W^2+m^2-M^2+\frac{4M^2}{1+\xi^t}+ O(1/W^2).
\label{Kin_quant_u}
\ee
and express $\xi^t$:
\be
\xi^t \simeq \frac{Q^2-t-M^2}{2W^2-Q^2+t-3 M^2}.
\ee

\setcounter{equation}{0}
\section{Charge conjugation  and $G$-parity issues}
\label{App_B}

\subsection{Notations and conventions}
\label{App_B1}

Below we employ our usual system notations of Ref.~\cite{Pire:2011xv}.
\begin{itemize}
\item Letters from the beginning of the Greek alphabet are reserved for the
${\rm SU}(2)$ isospin indices
$ \alpha,\,\beta,\,\gamma,\, \iota, \, \kappa  ={1,\,2 }$.
\item We have to distinguish between upper (contravariant) and lower (covariant) ${\rm SU}(2)$ isospin indices.
We introduce the totally antisymmetric tensor $\varepsilon_{\alpha \beta}$ for lowering indices and $\varepsilon^{\alpha \beta}$
for rising indices ($\varepsilon_{1 \,2}=\varepsilon^{1 \,2}=1$).
\item Letters from the middle of the Greek alphabet $\lambda$, $\mu$, $\nu$ denote the Lorentz indices.
\item Letters from the second half of the Greek alphabet $\rho, \, \tau,\, \chi$ are reserved for the Dirac indices.
\item Letters from the beginning of the Latin alphabet
$a,b,c\,...$
are reserved for indices of the adjoint representation of the ${\rm SU}(2)$ isospin group.
\item Letters $c_1$, $c_2$, $c_3$ stand for ${\rm SU}(3)$ color indices.
\end{itemize}

The nucleon field
$\bar{N}_\alpha$ ($N^\alpha$)
transforms  according to the covariant (contravariant) representation of the isospin ${\rm SU}(2)$.
 We adopt the following standard convention for the nucleon field
\cite{Itzykson}:
\be
&&
N^\alpha(x)=
\int \frac{d^3 k}{(2 \pi)^3}  \frac{M}{k_0}
\sum_{s=1,\, 2}
\big\{
e^{i k x} d^{\dag \, \alpha}(k,s) V(k,s)+e^{-i k x} b^{ \alpha}(k,s) U(k,s)
\big\} \,; \nonumber \\
&&
\bar{N}_\alpha(x)=
\int \frac{d^3 k}{(2 \pi)^3}  \frac{M}{k_0}
\sum_{s=1,\, 2}
\big\{
e^{i k x} {b^\dag}_\alpha (k,s) \bar{U}(k,s)+e^{-i k x} d_\alpha(k,s) \bar{V}(k,s)
\big\}\,.
\label{def_N_fiels}
\ee
Here spinors
$U(k,s)$
and
$\bar{U}(k,s) \equiv U^\dag (k,s) \gamma_0 $
describe a nucleon, respectively, in the initial and final states,
while spinors
$\bar{V}(k,s)\equiv V^\dag (k,s) \gamma_0$
and
$V(k,s)$
describe an antinucleon in the initial and final states.

The creation and annihilation operators in
(\ref{def_N_fiels})
satisfy the usual anticommutation relations for fermions 
\be
&&
\left\{
b^{ \alpha}(p,s), \, {b^\dag}_\beta(p',s')\right\}=(2 \pi)^3 \frac{p_0}{M} \delta^3(p-p') \delta_{ss'} \delta^{\alpha}_\beta\,;
\nonumber \\ &&
\left\{
d_\alpha(p,s), \,
{d}^{\dag \, \beta} (p',s')
\right\}
=(2 \pi)^3 \frac{p_0}{M} \delta^3(p-p') \delta_{ss'} \delta^\beta_\alpha \,.
\ee
The ``in'' nucleon state $|N_\alpha \rangle$ is defined according to:
\be
|N_1 \rangle \equiv | N_p(p,s) \rangle= {b}^{\dag}_1 (p,s) |0 \rangle \, ;  \ \ \  |N_2 \rangle \equiv  | N_n(p,s) \rangle= {b}^{\dag}_2 (p,s) |0 \rangle\,.
\label{nucleon_in_states}
\ee
Analogously,
the ``in'' antiparticle state  $|\bar{N}^\alpha \rangle$ is defined as:
\be
|\bar{N}^1 \rangle \equiv | N^{\bar{p}}(p,s) \rangle={d}^{\dag \,1 }  (p,s) |0 \rangle
 \, ;  \ \ \
|\bar{N}^2 \rangle \equiv | N^{\bar{n}}(p,s) \rangle={d}^{\dag \,2 }  (p,s) |0 \rangle\,.
\label{antinucleon_states}
\ee

The charge conjugation operator $\cal{C}$ has the following effect on the nucleon
creation and antinucleon annihilation operators \cite{Itzykson}
\be
&&
\mathcal{C} b^{ \iota} (p, \lambda) \mathcal{C}^\dag= \eta_N d_\iota (p, \lambda);
\label{Ceffect_N1}
\\ &&
\mathcal{C} d^{ \dag \, \iota} (p, \lambda) \mathcal{C}^\dag= \eta_N b^\dag_\iota (p, \lambda),
\label{Ceffect_N2}
\ee
where
$\eta_N$
denotes the nucleon field charge parity. These equalities are to be understood
not as equalities of ${\rm SU}(2)$ tensors but rather as equalities for their components.
From (\ref{Ceffect_N2}) and the hermitian conjugation of (\ref{Ceffect_N1}), we get, respectively:
\be
&&
\mathcal{C}|\bar{N}^\iota \rangle=  \eta_N |N_\iota \rangle;
\nonumber \\ &&
\mathcal{C}  |N_\iota \rangle = \eta_N^* |\bar{N}^\iota \rangle.
\label{CeffectstatesN}
\ee

\subsection{Antinucleon DA}
\label{App_B2}

The isotopic parametrization for the nucleon DA reads \cite{Pire:2011xv}
\be
&&
4\langle 0 | \varepsilon_{c_1 c_2 c_3} \Psi^{c_1 \alpha}_\rho(1) \Psi^{c_2 \beta}_\tau(2) \Psi^{c_3 \gamma}_\chi(3)| N_\iota(p) \rangle
\nonumber \\ &&
=
\varepsilon^{\alpha \beta} \delta^\gamma_\iota M^{N \{13\}}_{\rho \tau \chi}(1,2,3) +
\varepsilon^{\alpha \gamma} \delta^\beta_\iota M^{N \{12\}}_{\rho \tau \chi}(1,2,3),
\ee
where the invariant isospin amplitudes read
\be
&&
M^{N \{12\}}_{\rho \tau \chi}(1,2,3)= f_N
\left(
v_{\rho \tau, \chi}^N V^p(1,2,3)+ a_{\rho \tau, \chi}^N A^p(1,2,3) +
t_{\rho \tau, \chi}^N T^p(1,2,3)
\right); \nonumber \\ &&
M^{N \{13\}}_{\rho \tau \chi}(1,2,3)=M^{N \{12\}}_{\rho \chi \tau}(1,3,2).
\label{PVAT}
\ee
Here
$\{v^N, a^N, t^N\}_{\rho \tau, \, \chi} $
are the conventional Dirac structures,
\be
&&
v_{\rho \tau, \, \chi}^N=(\hat{p} C)_{\rho \tau} (\gamma^5 U(p))_\chi\,;
\ \
a_{\rho \tau, \, \chi}^N=(\hat{p} \gamma^5 C)_{\rho \tau} ( U(p))_\chi
\,;
\ \
t_{\rho \tau, \, \chi}^N=(\sigma_{p \mu}  C)_{\rho \tau} ( \gamma^\mu \gamma^5 U(p))_\chi,
\nonumber \\ &&
\label{DA structures}
\ee
where $C$ is the charge conjugation matrix;
$\sigma^{\mu \nu}= \frac{1}{2} [\gamma^\mu, \, \gamma^\nu]$
and
$\sigma^{p \nu} \equiv p_\mu \sigma^{\mu \nu}$.

The effect of the charge conjugation operator on the quark field
$\Psi^\alpha$
is \cite{Itzykson}:
\be
\mathcal{C} \Psi^\alpha \mathcal{C}^\dag= \eta_q C \bar{\Psi}^T_\alpha,
\label{Cpar_quarks}
\ee
where transposition refers to the Dirac index.
$C$ stands for the charge conjugation matrix and
$\eta_q$
is the corresponding charge parity.

Now using (\ref{Cpar_quarks}) together with (\ref{CeffectstatesN}),
we establish the link between nucleon and antinucleon DAs,
\be
&&
4\langle 0 | \varepsilon_{c_1 c_2 c_3} \Psi^{c_1 \alpha}_\rho(1) \Psi^{c_2 \beta}_\tau(2) \Psi^{c_3 \gamma}_\chi(3)| N_\iota(p) \rangle
\nonumber \\ &&
= 4\eta_N^* \langle 0| \varepsilon_{c_1 c_2 c_3}
\mathcal{C}^\dag \mathcal{C} \Psi^{c_1 \alpha}_\rho(1)
\mathcal{C}^\dag \mathcal{C} \Psi^{c_2 \beta}_\tau(2)
\mathcal{C}^\dag \mathcal{C} \Psi^{c_3 \gamma}_\chi(3) \mathcal{C}^\dag | \bar{N}^\iota(p) \rangle
\nonumber \\ &&
= 4 \eta_N^* \eta_q^3
\langle 0| \varepsilon_{c_1 c_2 c_3}
   \left( C\bar{\Psi}^{T \,c_1} _\alpha \right)_\rho
   \left( C\bar{\Psi}^{T \, c_2}_\beta \right)_\tau
   \left( C\bar{\Psi}^{T \, c_3}_\gamma \right)_\chi | \bar{N}^\iota(p) \rangle.
\ee

Now we can define the antinucleon DA as the ${\rm SU}(2)$ tensor:
\be
&&
4\langle 0 | \varepsilon_{c_1 c_2 c_3}
  \left(  \bar{\Psi}^{  \,c_1} _\alpha \right)_\rho
   \left(  \bar{\Psi}^{  \, c_2}_\beta \right)_\tau
   \left(  \bar{\Psi}^{  \, c_3}_\gamma \right)_\chi | \bar{N}^\iota(p) \rangle
=
\varepsilon_{\alpha \beta} \delta_\gamma^\iota M^{ \bar{N} \{13\}}_{\rho \tau \chi}(1,2,3) +
\varepsilon_{\alpha \gamma} \delta_\beta^\iota M^{\bar{N} \{12\}}_{\rho \tau \chi}(1,2,3),
\nonumber \\ &&
\ee
where
\be
&&
M^{\bar{N} \{12\}}_{\rho \tau \chi}(1,2,3)=
f_N
\left(
v_{\rho \tau, \chi}^{\bar{N}} V^{\bar{p}}(1,2,3)+ a_{\rho \tau, \chi}^{\bar{N}} A^{\bar{p}}(1,2,3) +
t_{\rho \tau, \chi}^{\bar{N}} T^{\bar{p}}(1,2,3)
\right);
\nonumber \\ &&
M^{\bar{N} \{13\}}_{\rho \tau \chi}(1,2,3)=M^{\bar{N} \{12\}}_{\rho \chi \tau}(1,3,2),
\label{AntiPVAT}
\ee
where the relevant Dirac structures are
\be
&&
(v_{\rho \tau, \chi}^{\bar{N}})^T \equiv (C^\dag)_{\rho \rho'} (C^\dag)_{\tau \tau'} (C^\dag)_{\chi \chi'} v_{\rho' \tau', \chi'}^{N}=
(C \hat{p})^T_{\rho \tau} (\bar{V} \gamma^5)^T_\chi;
\nonumber \\ &&
(a_{\rho \tau, \chi}^{\bar{N}})^T \equiv(C^\dag)_{\rho \rho'} (C^\dag)_{\tau \tau'} (C^\dag)_{\chi \chi'} a_{\rho' \tau', \chi'}^{N}=
(C \gamma_5 \hat{p})^T_{\rho \tau}  (\bar{V} )^T_\chi
\nonumber \\ &&
(t_{\rho \tau, \chi}^{\bar{N}})^T \equiv(C^\dag)_{\rho \rho'} (C^\dag)_{\tau \tau'} (C^\dag)_{\chi \chi'} t_{\rho' \tau', \chi'}^{N}=
(C \sigma_{p \mu})^T_{\rho \tau}  (\bar{V} \gamma^\mu \gamma_5)^T_\chi
\label{DiracStrBarNT}
\ee
Lifting the transposition with respect to the Dirac indices in
(\ref{DiracStrBarNT})
one gets
\be
&&
v_{\rho \tau, \chi}^{\bar{N}}= (C \hat{p})_{\rho \tau} (\bar{V} \gamma^5)_\chi; \nonumber \\ &&
a_{\rho \tau, \chi}^{\bar{N}}=(C \hat{p} \gamma_5 )_{\rho \tau}  (\bar{V} )_\chi; \nonumber \\ && 
t_{\rho \tau, \chi}^{\bar{N}}=(C \sigma_{p \mu})_{\rho \tau}  (\bar{V} \gamma^\mu \gamma_5)_\chi.
\ee

Therefore, we conclude that as the consequence of charge conjugation invariance antiproton DAs occurring in
(\ref{AntiPVAT}) are expressed as follows through the usual proton DAs of (\ref{PVAT}):
\be
\{V, \, A, \, T\}^{\bar{p}}(y_1,y_2,y_3)=\frac{1}{\eta_N^* \eta_q^3} \{V, \, A, \, T\}^{p}(y_1,y_2,y_3).
\ee

\subsection{Antinucleon to pion TDAs}
\label{App_B3}
Below we establish the relations between the antinucleon to pion and nucleon to pion TDAs.
Our main line of reasoning is analogous to that in the case of the antinucleon DAs.
However, charged pions are not eigenstates of the charge conjugation operator $\cal C$.
To work out the isotopic formalism for the antinucleon to pion TDAs it is convenient
to employ the concept of $G$-parity (see {\it e.g.} \cite{EricsonWeise}), which is the combination of the charge conjugation and the
rotation around the second axis of the isospin space,
\be
{\cal G}= {\cal C} e^{i \pi I_2}.
\ee

First, let us work out the effect of $\cal G$ on the quark and antiquark fields $\Psi^\alpha$,
\be
&&
{\cal G}  \Psi^\alpha {\cal G}^\dag= {\cal C} e^{i \pi I_2} \Psi^\alpha e^{-i \pi I_2} {\cal C}^\dag.
\ee
According to the conventions of \cite{Pire:2011xv},
we choose to transform
$\bar{\Psi}_\alpha$
field  according to the covariant representation and to transform
$\Psi^\alpha$
field  according to the contravariant representation of the isospin $  {\rm SU} (2)$:
\be
\left[
I_a, \bar{\Psi}_\alpha
\right]
= \frac{1}{2} \left( \sigma_a \right)^\beta_{\; \alpha} \bar{\Psi}_\beta\,;
\ \ \
\left[
I_a, \Psi^\alpha\right]=- \frac{1}{2} \left( \sigma_a \right)^\beta_{\; \alpha} \Psi^\alpha\,,
\label{isospin_commutators_with_N}
\ee
where $I_a$ stand for the group generators and
$\sigma_a$
are the Pauli matrices.

The effect of the isospin rotation around the second axis then reads:
\be
e^{i \pi I_2} (\Psi^\alpha)_\rho e^{-i \pi I_2}=  \left( \cos \frac{\pi}{2} \delta^\alpha_{\;  \beta} + \sin \frac{\pi}{2} (-i \sigma_2)^\alpha_{\; \beta}  \right) (\Psi^\beta)_\rho.
\ee
Thus, we conclude that
\be
&&
{\cal G} u_\rho {\cal G}^\dag= \eta_q (-1) (C\bar{d})^T_\rho; \ \ \ {\cal G} d_\rho {\cal G}^\dag= \eta_q   (C\bar{u})^T_\rho.
\label{transfud}
\ee

We  also need to specify the effect of $\cal G$ on pion ``out'' states and nucleon ``in'' states.
The triplet of pions has negative $G$ parity,
\be
\langle \pi | {\cal G}^\dag= - \langle \pi |.
\label{transpi}
\ee

Employing our isospin conventions for the nucleon states: summarized in   Appendix~\ref{App_B1}
we conclude that:
\be
{\cal G} |N_p \rangle= \eta_N^* (-1) | \bar{N}^{\bar{n}} \rangle; \ \ \
{\cal G} |N_n \rangle= \eta_N^*  | \bar{N}^{\bar{p}} \rangle.
\label{transpn}
\ee

For antinucleon to pion TDAs we can write down the following
parametrization
\be
&&
4\langle  \pi_a | \hat{\bar{O}}_{\alpha \beta \gamma \; \rho \tau \chi}(1,\,2,\,3) | \bar{N}^\iota \rangle
\nonumber \\ &&
= (f_a)_{\{\alpha \beta \gamma \}}^{\iota}
M^{(\pi \bar{N})_{3/2}}_{\rho \tau \chi}
(1,2,3)
+
\varepsilon_{\alpha \beta} (\sigma_a)^\iota_{\ \gamma}  M^{(\pi \bar{N})_{1/2} \, \{1 3 \}}_{\rho \tau \chi}
(1,2,3)
\nonumber \\ &&
+\varepsilon_{\alpha \gamma} (\sigma_a)^\iota_{\ \beta} M^{(\pi \bar{N})_{1/2} \, \{1 2 \}}_{\rho \tau \chi}
(1,2,3)\,,
\label{BarNpi_isospin_dec}
\ee
where the tensor $(\bar{f}_a)_{\{\alpha \beta \gamma\}}^{  \iota}$, which is totally symmetric in 
$\alpha$, $\beta$, $\gamma$, 
reads:
\be
&&
(\bar{f}_a)_{\{\alpha \beta \gamma\}}^{  \iota}
=
\frac{1}{3}
\left(
(\sigma_a^T)_{\alpha}^{\; \; \delta} \, \varepsilon_{\delta \beta} \, \delta^{\iota}_{\; \; \gamma}
+
(\sigma_a^T)_{\alpha}^{\; \delta} \varepsilon_{\delta \gamma} \delta^{\iota}_{\; \beta}+
(\sigma_a^T)_{\beta}^{\; \delta} \varepsilon_{\delta \gamma} \delta^{\iota}_{\; \alpha}
\right)\,;
\nonumber \\ && {\rm  since} \ \ \
(\sigma_a^T)_{\alpha}^{\; \delta}=(\sigma_a)_\alpha^{\; \; \delta} \equiv \varepsilon_{\alpha \kappa}\, (\sigma_a)^\kappa_{\; \; \theta} \, \varepsilon^{\theta \delta}
\label{Def_ta_tensor}
\ee

The isospin and permutation symmetry identities for the  isospin-$\frac{1}{2}$
and isospin-$\frac{3}{2}$ invariant  $\pi \bar{N}$ TDAs
$ M^{(\pi \bar{N})_{1/2} \, \{1 2 \}}$,
$ M^{(\pi \bar{N})_{1/2} \, \{1 3 \}}$
and
$ M^{(\pi \bar{N})_{3/2}}$
are the same as those for the relevant $\pi N$ TDAs
(see Sec.~4 of Ref.~\cite{Pire:2011xv}). In particular,
\be
M^{(\pi \bar{N})_{1/2} \, \{1 3 \}}_{\rho \tau \chi}(1,2,3)=
M^{(\pi \bar{N})_{1/2} \, \{1 2 \}}_{\rho  \chi \tau}(1,3,2).
\ee

Now,
using $G$-parity operator $\cal G$ instead of $\mathcal{C}$,
we repeat the derivation of App.~\ref{App_B2} and establish
the link between $\pi N$ and $\pi \bar{N}$ TDAs.

For example,
\be
&&
 -\frac{\sqrt{2}}{3} M^{(\pi N)_{3/2}}(1,2,3)+ \sqrt{2}  M^{(\pi N)_{1/2}}(1,2,3)
 \nonumber \\ &&=
\langle \pi^-| {\cal G}^\dag {\cal G} u_\rho (1) {\cal G}^\dag {\cal G} u_\tau(2)
{\cal G}^\dag {\cal G} d_\chi(3) {\cal G}^\dag {\cal G} | N_n \rangle \nonumber \\ &&
=(-1)     \eta_q^3 \eta_N^*
 (C)_{\rho \rho'} (C)_{\tau \tau'} (C)_{\chi \chi'}
 \langle \pi^-| (\bar{d}_{\rho'})^T (1)  (\bar{d}_{\tau'})^T (2)  (\bar{u}_{\chi'})^T(3) | \bar{N}^{\bar{p}} \rangle
  \nonumber \\ &&=
  \eta_q^3 \eta_N^* (C)_{\rho \rho'} (C)_{\tau \tau'} (C)_{\chi \chi'}
    \nonumber \\ && \times
  \big\{
  -\frac{\sqrt{2}}{3}  \left( M^{(\pi \bar{N})_{3/2}}_{\rho' \tau' \chi'}(1,2,3) \right)^T+ \sqrt{2}
   \left( M^{(\pi \bar{N})_{1/2}}_{\rho' \tau' \chi'}(1,2,3) \right)^T
  \big\}.
\ee

One may check that the general parametrization for $\pi \bar{N}$ TDAs (\ref{BarNpi_isospin_dec})
is consistent with that for $\pi N$ TDAs (see eq.~(54) of Ref.~\cite{Pire:2011xv})
once
 \be
 &&
  \left( M^{(\pi \bar{N})_{1/2}}_{\rho \tau \chi}(1,2,3) \right)^T = \frac{1}{\eta_N^* \eta_q^3}
  (C^\dag)_{\rho \rho'} (C^\dag)_{\tau \tau'} (C^\dag)_{\chi \chi'}
  M^{(\pi N)_{1/2}}_{\rho' \tau' \chi'}(1,2,3);
\nonumber \\ &&
 \left( M^{(\pi \bar{N})_{3/2}}_{\rho \tau \chi}(1,2,3) \right)^T = \frac{1}{\eta_N^* \eta_q^3}
  (C^\dag)_{\rho \rho'} (C^\dag)_{\tau \tau'} (C^\dag)_{\chi \chi'}
  M^{(\pi N)_{3/2}}_{\rho' \tau' \chi'}(1,2,3),
 \ee
where the transposition refers to the Dirac indices.
The Dirac structures
$s^{\pi \bar{N}} \equiv \{v^{\pi \bar{N}}_{1,2}, \, a^{\pi \bar{N}}_{1,2}, \, t^{\pi \bar{N}}_{1,2,3,4} \} $ occurring in the parametrization of
the isospin-$\frac{1}{2}$ and isospin-$\frac{3}{2}$ invariant amplitudes
$M^{(\pi \bar{N})_{1/2}}$ and $M^{(\pi \bar{N})_{3/2}}$
are defined by
\be
\left(
s^{\pi \bar{N}}_{\rho \tau, \chi}
\right)^T= (C^\dag)_{\rho \rho'} (C^\dag)_{\tau \tau'} (C^\dag)_{\chi \chi'} s_{\rho' \tau', \chi'}^{ \pi N},
\ee
where
$s_{\rho' \tau', \chi'}^{ \pi N}$
are the Dirac structures occurring in the parametrization of $\pi N$ TDAs
(see eq.~(12) of Ref.~\cite{Pire:2011xv}).
For the relevant $\pi \bar{N}$  TDAs
(\ref{TDAsScalar})
we, thus, get
\be
H^{\pi \bar{N}}(x_1,x_2,x_3, \xi, \Delta^2)= \frac{1}{\eta_N^* \eta_q^3} H^{\pi N}(x_1,x_2,x_3, \xi, \Delta^2).
\ee

\setcounter{equation}{0}
\section{Amplitude calculation within the $u$-channel factorization mechanism}
\label{App_C}

Below we argue that the amplitude of
(\ref{BarNNannihilation reaction})
within the $t$-channel factorization mechanism can be obtained from that within the
$u$-channel factorization mechanism  (see Sec.~\ref{Sec-2})
with the obvious change of the kinematical variables:
\be
&&
p_N  \rightarrow p_{\bar{N}};  \ \ \   p_{\bar{N}}  \rightarrow p_{{N}}; \nonumber \\ &&
\Delta^u \rightarrow \Delta^t \ \ \ (u  \rightarrow t); \nonumber \\ &&
\xi^u  \rightarrow \xi^t.
\label{tu_change}
\ee
Since isotopic invariance plays no particular role in the present consideration
we consider one flavor of quarks and one ``pion'' with negative $C$-parity: $\eta_\pi=-1$.
Generalization for the physical case is straightforward with the use
of the $G$-parity formalism described in App.~\ref{App_B3}.

We introduce the following parametrization for the corresponding matrix elements:
\be
&&
\langle \pi | \Psi_\rho(x_1) \Psi_\tau(x_2) \Psi_\chi(x_3) | N(p_N) \rangle= A_{\rho \tau}(p_N,\Delta^u) (B U(p_N))_\chi
H(x_i, \xi^u, u)
\nonumber \\ &&
\langle 0 | \bar{\Psi}_{\bar{\rho}}^T(y_1)  \bar{\Psi}_{\bar{\tau}}^T(y_2)  \bar{\Psi}_{\bar{\chi}}^T(y_3) | \bar{N} (p_{\bar{N}}) \rangle=
D_{\bar{\rho} \bar{\tau}}(p_{\bar{N}}) (\bar{V}(p_{\bar{N}})E)_{\bar{\chi}} \Phi(y_i),
\label{DiracStr_tchannel}
\ee
where
$A(p_N,\Delta_u)$, $B$, $D(p_{\bar{N}})$ and $E$ are the relevant Dirac structures;
$H(x_i, \xi^u, u)$
and
$\Phi(y_i)$
denote the scalar TDAs and DAs respectively.

We now express the matrix elements occurring in the $t$-channel factorization scheme through the Dirac structures
of the $u$-channel factorization scheme occurring in (\ref{DiracStr_tchannel}):
\be
&&
\langle \pi |
\bar{\Psi}^T_{\rho}(x_1)
\bar{\Psi}^T_{\tau}(x_2)
\bar{\Psi}^T_{\chi}(x_3)
| \bar{N} (p_{\bar{N}}) \rangle
\nonumber \\ &&
=\frac{1}{{\eta_N^*}  {\eta_q}^3  \eta_\pi
}
C_{\rho \rho'}^\dagger  C_{\tau \tau'}^\dagger  C_{\chi \chi'}^\dagger \langle \pi |
\Psi_{\rho'}(x_1)
\Psi_{\tau'}(x_2)
\Psi_{\chi'}(x_3)
|  {N} (p_{\bar{N}}) \rangle
\nonumber \\ &&
= \frac{1}{{\eta_N^*} {\eta_q}^3 \eta_\pi}
(C^\dag A {C^\dag}^T)_{\rho \tau} 
(C^\dag BU)_\chi H(x_i, \xi^t, t);
\ee
\be
&&
\langle 0|  \Psi_{\bar{\rho}}(y_1)
\Psi_{\bar{\tau}}(y_2)
\Psi_{\bar{\chi}}(y_3)|N(p_N) \rangle
\nonumber \\ &&
=\frac{1}{{\eta_N} {\eta_q^*}^3 }
C_{\bar{\rho} \bar{\rho}'}
C_{\bar{\tau} \bar{\tau}'}
C_{\bar{\chi} \bar{\chi}'}
\langle 0|
\bar{\Psi}_{\bar{\rho}'}^T(y_1)
\bar{\Psi}_{\bar{\tau}'}^T(y_2)
\bar{\Psi}_{\bar{\chi}'}^T(y_2)| \bar{N}(p_N) \rangle
\nonumber \\ &&
=\frac{1}{{\eta_N} {\eta_q^*}^3 }
(C D C^T)_{\bar{\rho} \bar{\tau}} 
 \left(\bar{V} E C^T \right)_{\bar{\chi}} \Phi(y_i).
\ee
In this way we express antinucleon to pion TDAs  (respectively, nucleon DAs) through
nucleon to pion TDAs (respectively, antinucleon DAs).

\begin{figure}[h]
 \begin{center}
 \epsfig{figure=  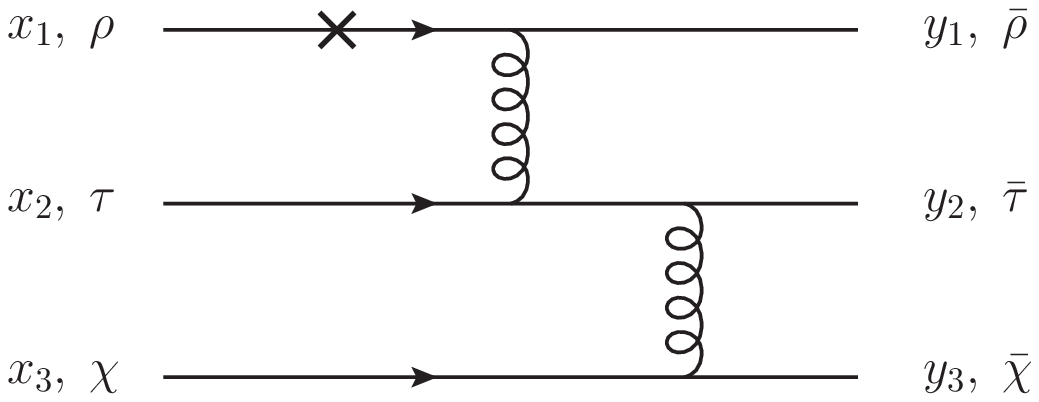 , height=2.5cm} \ \ \
  \epsfig{figure= 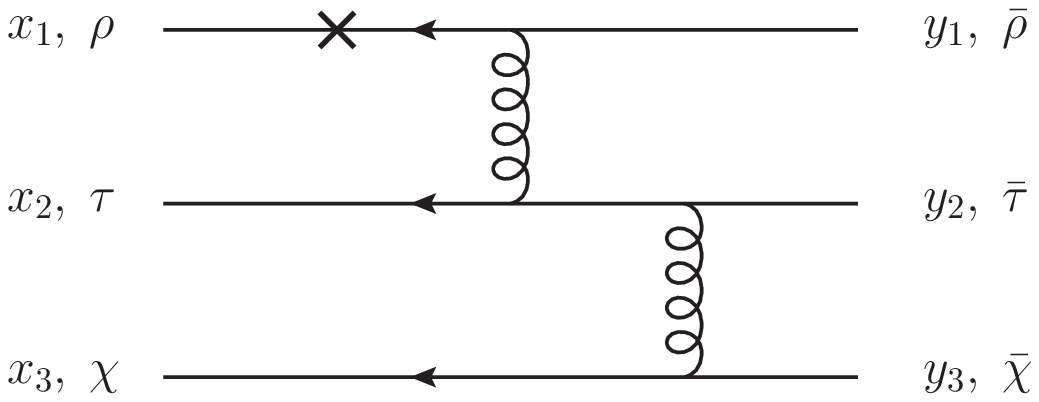 , height=2.5cm}
  \end{center}
     \caption{ Typical graphs  corresponding to $N \bar{N} \to \gamma^*(q) M(p_M)$. {\bf Left panel:}   $u$-channel regime.
           {\bf Right panel:} $t$-channel regime. The crosses represent the virtual photon vertex.
      }
\label{FigAppB}

\end{figure}

Now we proceed with the generic contributions to the amplitude.
Each of three lines of the $u$-channel amplitude graph depicted in Fig.~\ref{FigAppB}
corresponds to a matrix in the Dirac space. We denote these matrices as
$(L_1)_{\bar{\rho} \rho}$,
$(L_2)_{\bar{\tau} \tau}$
and
$(L_3)_{\bar{\chi} \chi}$.
Up to the obvious changes due to
$p_N \leftrightarrow p_{\bar{N}}$,
the matrices corresponding to three lines of the
$t$-channel graph are
$(L_1^T)_{\bar{\rho} \rho}$,
$(L_2^T)_{\bar{\tau} \tau}$
and
$(L_3^T)_{\bar{\chi} \chi}$.

With these notations the spinor structure of the amplitude corresponding to the $u$-channel graph reads
\be
S^u=D_{\bar{\rho} \bar{\tau}}(p_{\bar{N}}) (\bar{V}(p_{\bar{N}}) E)_{\bar{\chi}}
(L_1)_{\bar{\rho} \rho}
(L_2)_{\bar{\tau} \tau}
(L_3)_{\bar{\chi} \chi}
A_{\rho \tau}(p_N, \Delta^u) (BU(p_N))_\chi.
\ee
The spinor structure of the amplitude corresponding to the $t$-channel graph then reads:
\be
&&
S^t \nonumber \\ &&
= \frac{(-1)^3}{{\eta_\pi} }D_{ {\rho}  {\tau}}(p_{ {N}}) (\bar{V}(p_{ {N}}) E)_{ {\chi}}
(C^\dag L_1^T C)_{{\rho} \bar{\rho}}
(C^\dag L_2^T C)_{{\tau} \bar{\tau}}
(C^\dag L_3^T C)_{ {\chi} \bar{\chi}}
A_{\bar{\rho} \bar{\tau}}(p_{\bar{N}}, \Delta^u) (BU(p_{\bar{N}}))_{\bar{\chi}}.
\nonumber \\ &&
\label{Ru}
\ee
Given that for any of $21$ relevant graphs presented in Table~I of Ref.~\cite{Lansberg:2007ec}
$L_1$, $L_2$ and $L_3$
contain an odd number of the Dirac matrices $\gamma_\mu$ (and do not contain $\gamma_5$)
we conclude that
$S^u$ turns into $\eta_\pi^{-1} S^t$ 
 with the change
(\ref{tu_change}).
Thus we arrive at the conclusion that the $C$-invariance results in a symmetry between the $u$- and $t$-channel  factorization
mechanisms. The amplitudes for $N \bar{N} \to \gamma^* \pi$
within the two mechanisms have the forms,
\be
&&
\mathcal{M}^{\lambda}_{s_N s_{\bar{N}}}=
\mathcal{C} \frac{1}{Q^4}
\Big[
\bar{V}(p_{\bar{N}},s_{\bar{N}}) \hat{\epsilon}^*(\lambda) \gamma_5 U(p_N,s_N)
\mathcal{I}(\xi^u, u)
\nonumber \\ &&
-
\frac{1}{M}
\bar{V}(p_{\bar{N}},s_{\bar{N}}) \hat{\epsilon}^*(\lambda) \hat{\Delta}_T^u \gamma_5 U(p_N,s_N)
\mathcal{I}'(\xi^u, u)
\Big];
\label{Ampl_u_regime}
\ee
\be
&&
\mathcal{M}^{\lambda}_{s_N s_{\bar{N}}}=\frac{
1
}{{\eta_\pi} }
\mathcal{C} \frac{1}{Q^4}
\Big[
\bar{V}(p_{ N},s_{N}) \hat{\epsilon}^*(\lambda) \gamma_5 U(p_{\bar{N}},s_{\bar{N}})
\mathcal{I}(\xi^t, t)
\nonumber \\ &&
-
\frac{1}{M}
\bar{V}(p_N,s_N) \hat{\epsilon}^*(\lambda) \hat{\Delta}_T^t \gamma_5 U(p_{\bar{N}},s_{\bar{N}})
\mathcal{I}'(\xi^t, t)
\Big],
\label{Ampl_t_regime}
\ee
with the same functions
$\cal{I}$ and $\cal{I}'$
(\ref{Def_IandIprime})
standing both in
(\ref{Ampl_t_regime})
and
(\ref{Ampl_u_regime}).
Although a somewhat unusual choice of the Dirac spinors is employed for $S^t$ in (\ref{Ampl_u_regime}),
this does not lead to any  particular complications for the calculation of unpolarized cross sections.
Indeed, (\ref{tu_change})
can be used at the cross section level after the summation over nucleons spins is performed.
As a result, the forward and the backward peaks of the unpolarized cross section  possess the
identical structure.

\end{document}